\documentclass[]{iopart}
\usepackage{graphicx} 

\expandafter\let\csname equation*\endcsname\relax
\expandafter\let\csname endequation*\endcsname\relax

\usepackage{amsmath,amsfonts,amssymb}
\usepackage[caption=false]{subfig}
\usepackage[dvipsnames,usenames]{xcolor}
\usepackage{fancyhdr}
\usepackage[normalem]{ulem}

\usepackage{censor}
\StopCensoring

\usepackage{etoolbox}
\newbool{showchanges}
\boolfalse{showchanges}

\usepackage{etoolbox}
\usepackage{xspace}
\definecolor{linkcolor}{rgb}{0.0,0.3,0.5}
\usepackage{footmisc}
\usepackage[
unicode, 
colorlinks=true,
linkcolor=linkcolor,
citecolor=linkcolor,
filecolor=linkcolor,
urlcolor=linkcolor,
hyperfootnotes=true,
]{hyperref}
\usepackage[all]{hypcap}
\usepackage[T1]{fontenc}
\usepackage{pifont}
\usepackage{mathtools}
\usepackage[numbers,square,sort&compress]{natbib}
\usepackage[utf8]{inputenc}
\usepackage{enumitem}
\usepackage{standalone}
\usepackage{tikz}
\usetikzlibrary{decorations.pathmorphing}
\usetikzlibrary{shapes.geometric}

\newcommand\centerofmass{%
    \tikz[radius=0.4em,scale=0.5] {%
        \fill (0,0) -- ++(0.4em,0) arc [start angle=0,end angle=90] -- 
++(0,-0.8em) arc [start angle=270, end angle=180];%
        \draw (0,0) circle;%
    }%
}

\newcommand{\thetitle}{Lessons for adaptive mesh refinement in numerical 
relativity}
\newcommand{\grchombo}{\textsc{GRChombo}\xspace}
\newcommand{\chombo}{\textsc{Chombo}\xspace}
\newcommand{\lean}{\textsc{Lean}\xspace}
\newcommand{\iu}{\mathrm{i}\mkern1mu}
\newcommand{\lmax}{l_{\max}}

\ifbool{showchanges}{
    \newcommand{\change}[2]{\old{#1} \new{#2}}
}{
    \newcommand{\change}[2]{#2}
}

\makeatletter
\newcommand{\mainmatter}{%
  \setcounter{footnote}{0}%
  \patchcmd{\@makefntext}{\fnsymbol}{\arabic}{}{}%
  \patchcmd{\@thefnmark}{\fnsymbol}{\arabic}{}{}%
  \def\@makefnmark{\textsuperscript{\arabic{footnote}}}%
}
\makeatother

\begin{document}
\newcounter{count}

\pagestyle{fancy}
\lhead{\thetitle}
\chead{}
\rhead{\thepage}
\lfoot{}
\cfoot{}
\rfoot{}

\hfill{\scriptsize KCL-PH-TH/2021-89}

\begin{center}
\title{\large\thetitle}
\end{center}

\author{
\censor{Miren Radia}\texorpdfstring{$^{1}$}{},
\censor{Ulrich Sperhake}\texorpdfstring{$^{1,2,3}$}{},
\censor{Amelia Drew}\texorpdfstring{$^{1}$}{},
\censor{Katy Clough}\texorpdfstring{$^{4}$}{},
\censor{Pau Figueras}\texorpdfstring{$^{4}$}{},
\censor{Eugene A. Lim}\texorpdfstring{$^{5}$}{},
\censor{Justin L. Ripley}\texorpdfstring{$^{1}$}{},
\censor{Josu C. Aurrekoetxea}\texorpdfstring{$^{6}$}{},
\censor{Tiago Fran\c{c}a}\texorpdfstring{$^{4}$}{},
\censor{Thomas Helfer}\texorpdfstring{$^{2}$}{},
}

\address{$^{1}$~\xblackout{
Department of Applied Mathematics and Theoretical Physics, 
Centre for Mathematical Sciences, University of Cambridge, Wilberforce Road, 
Cambridge CB3 0WA, United Kingdom}}

\address{$^{2}$~\xblackout{
Department of Physics and Astronomy, Johns Hopkins University, 
3400 N. Charles Street, Baltimore, Maryland 21218, USA}}

\address{$^{3}$~\xblackout{
California Institute of Technology, Pasadena, California 91125, 
USA}}

\address{$^{4}$~\xblackout{
Geometry, Analysis and Gravitation, School of Mathematical 
Sciences, Queen Mary University of London, Mile End Road, London E1 4NS, 
United Kingdom}}

\address{$^{5}$~\xblackout{
Theoretical Particle Physics and Cosmology Group, Physics 
Department, Kings College London, Strand, London WC2R 2LS, United Kingdom}}

\address{$^{6}$~\xblackout{
Astrophysics, Denys Wilkinson Building, University of Oxford, 
Keble Road, Oxford OX1 3RH, United Kingdom}}

\ead{\xblackout{m.r.radia@damtp.cam.ac.uk,
     u.sperhake@damtp.cam.ac.uk,
     a.drew@damtp.cam.ac.uk,
     k.clough@qmul.ac.uk,
     p.figueras@qmul.ac.uk,
     eugene.a.lim@gmail.com,
     lloydripley@gmail.com,
     josu.aurrekoetxea@physics.ox.ac.uk,
     t.e.franca@qmul.ac.uk,
     thomashelfer@live.de}
    }

\begin{abstract}
We demonstrate the flexibility and utility of the Berger-Rigoutsos Adaptive 
Mesh Refinement (AMR) algorithm used in the open-source numerical relativity 
code \grchombo for generating gravitational waveforms from binary black-hole 
inspirals, and for studying other problems involving non-trivial matter 
configurations. We show that \grchombo can produce high quality binary 
black-hole waveforms through a code comparison with the established numerical 
relativity code \lean. We also discuss some of the technical challenges 
involved in making use of full AMR (as opposed to, e.g. moving box mesh 
refinement), including the numerical effects caused by using various refinement 
criteria when regridding. We suggest several ``rules of thumb'' for when to use 
different tagging criteria for simulating a variety of physical phenomena. We 
demonstrate the use of these different criteria through example evolutions of a 
scalar field theory. Finally, we also review the current status and general 
capabilities of \grchombo.
\end{abstract}

\begin{indented}
\item[]\textit{Keywords\/}: 
numerical relativity, 
adaptive mesh refinement, gravitational waves, compact objects, 
computational methods
\end{indented}

\maketitle
\mainmatter

\section{Introduction}

One of the key theoretical achievements that underpins the momentous detection 
of gravitational waves (GWs) from the inspiralling black-hole (BH) binary 
GW150914 \cite{LIGOScientific:2016aoc} is the the numerical relativity (NR) 
breakthrough in binary BH modeling in 2005 \cite{Pretorius:2005gq, 
Campanelli:2005dd, Baker:2005vv}. About 90 compact binary merger events have by 
now been observed by the GW detector network 
\cite{LIGOScientific:2018mvr,LIGOScientific:2020ibl,LIGOScientific:2021djp} 
with a wide range of total masses and mass ratios of the binary constituents. 
Indeed, it is a generic feature of general relativity and its character as a 
highly non-linear theory that its solutions often span across a large range of 
spatial and temporal scales. Combined with the inherent limits of computational 
resources, it follows that any finite difference numerical code will require 
some form of spatial and temporal mesh refinement to fully capture the dynamics 
of these solutions.

In the numerical relativity (NR) community, many codes rely on the technique of 
so-called ``moving boxes'' for mesh refinement, where a hierarchy of nested 
boxes with increasingly fine meshes is centered around specified points (also 
sometimes referred to as the ``box-in-box'' approach). Within this framework, 
boxes move around either dynamically or along predetermined paths, in order to 
track objects' \change{predicted }{}trajectories\footnote{In addition, boxes may be 
allowed to merge if they come close enough together. These codes include those 
built upon the popular \textsc{Cactus} computational framework 
\cite{Goodale2002a, Loffler:2011ay,Schnetter:2003rb} such as the 
\textsc{McLachlan} \cite{Brown:2008sb}, \textsc{LazEv} \cite{Zlochower:2005bj}, 
\textsc{Maya} \cite{Herrmann:2007zz}, \lean 
\cite{Sperhake:2006cy,Zilhao:2010sr} and \textsc{Canuda} \cite{Witek:2021can} 
codes, and the BAM code 
\cite{Bruegmann:1996kz,Bruegmann:2006ulg,Thierfelder:2011yi,Marronetti:2007ya}.
\label{fn:moving-box-codes}
}. This technique has proved remarkably successful, 
particularly in the case of generating gravitational waveforms from binaries of 
compact objects for the template banks for gravitational wave (GW) detectors, 
such as LIGO-Virgo-KAGRA 
\cite{KAGRA:2013rdx,Szilagyi:2014fna,Kumar:2013gwa,Boyle:2019kee, 
Szilagyi:2015rwa}. Moving box codes have matured to allow exploration of a wide 
variety of physics with a plethora of diagnostic tools 
\cite{Alcubierre:2008co,Baumgarte:2010bk}. 

However, there are classes of problems for which the moving boxes technique 
becomes impractical due to the topology of the system. Here, the use of ``fully 
adaptive'' mesh refinement (AMR) is required where the mesh dynamically adjusts 
itself in response to the underlying physical system being simulated, following 
user-specified mesh refinement criteria. In general, there are two broad 
classes of AMR, depending on whether newly refined meshes are added to the grid 
on a cell-by-cell ``tree-structured'' basis 
\cite{Kravtsov:1997vm,Teyssier:2001cp,Barrera-Hinojosa:2019mzo} or on a 
box-by-box ``block-structured'' basis. In this work, we will exclusively 
discuss the latter.

In block-structured AMR, first described and implemented by 
Berger \textit{et al.}~\cite{Berger:1984zza}, the computational domain is built 
from a hierarchy of increasingly fine levels, with each one containing a set of 
(not necessarily contiguous) boxes of meshes, with the only condition being that 
a finer mesh must lie on top of one \emph{or possibly more} meshes from the 
next coarsest level. It is important to stress that this means it is allowed for 
a fine mesh to straddle more than one coarse mesh---in other words the grid 
structure is \emph{level}-centric rather than box-centric. In contrast to the 
moving boxes approach, this approach allows for highly flexible 
``many-boxes-in-many-boxes'' mesh topologies, enabling the study of dynamical 
systems where the spacetime dynamics are not driven by localized compact systems 
e.g. in studying non spherical collapse scenarios \cite{East:2019bwu, 
Pretorius:2018lfb}, higher dimensional black holes/black string evolution 
\cite{Lehner:2010pn,Figueras:2020dzx,Andrade:2020dgc,Bantilan:2019bvf, 
Figueras:2017zwa,Figueras:2015hkb}, cosmic string evolution 
\cite{Drew:2019mzc,Helfer:2018qgv,Aurrekoetxea:2020tuw}, and the behavior of 
strongly inhomogeneous cosmological spacetimes \cite{East:2015ggf, 
Cook:2020oaj, Ijjas:2020dws,Ijjas:2021gkf,Joana:2020rxm, 
Aurrekoetxea:2019fhr,Clough:2017efm,Clough:2016ymm,deJong:2021bbo}.

Despite its advantages, AMR is a double-edged sword and its flexibility comes 
with a cost---each coarse-fine transition may introduce unwanted interpolation 
and prolongation errors whose magnitude depends on the order of the coarse-fine 
boundary operators, in addition to introducing a “hard surface” which can 
generate spurious unphysical reflections. We emphasize that AMR should not be 
treated as a “black box”, but requires careful control and fine-tuning of 
refinement criteria, that often depend on the physics being simulated, in order 
to work effectively. In particular, the creation/destruction of a finer grid is 
determined  by the tagging of cells for refinement, which in turn is controlled 
by a \emph{tagging criterion}. Although this ability to refine regions can be 
incredibly powerful, in practice it can be difficult to manage the exact 
placement of refined grids. Furthermore, we find that the management of 
coarse-fine boundaries in dynamically sensitive regions of spacetime, such as 
near apparent horizons, is essential for producing accurate results. 

In this paper we explain some of the tagging criteria and numerical techniques 
we have used to obtain convergent, reliable results when using block-structured 
AMR. We will discuss these issues in the context of the AMR NR code \grchombo 
\cite{Andrade:2021,Clough:2015sqa}\footnote{Adaptive mesh refinement is now 
being used in several other NR codes. For example, those based on the PAMR/AMRD 
mesh refinement libraries \cite{Pretorius:2004jg,Pretorius:2005ua,East:2011aa}, 
the \textsc{Hahndol} code \cite{Baker:2005vv}, which uses PARAMESH 
\cite{MacNeice2000}, \change{}{the \textsc{HAD} code \cite{Anderson:2006ay}, }
and the  pseudospectral codes, \textsc{SpEC} \cite{Pfeiffer:2002wt}, 
\textsc{bamps} \cite{Hilditch:2015aba} and SFINGE 
\cite{Meringolo:2020jsu} in which the AMR implementation is somewhat different 
to finite difference codes like \grchombo. More recently, \textsc{CosmoGRaPH} 
\cite{Mertens:2015ttp}, and \textsc{Simflowny} \cite{Palenzuela:2018sly} both 
based on the SAMRAI library \cite{Hornung:2002,Gunney:2016,SAMRAIWebsite}, 
\textsc{GRAthena++} \cite{Daszuta:2021ecf}, \change{}{\textsc{Gmunu} 
\cite{Cheong:2020kpv} and \textsc{Dendro-GR}
\cite{Fernando:2018mov}, all} based on \change{an}{} oct-tree AMR, and 
GRAMSES \cite{Barrera-Hinojosa:2019mzo} have been introduced. Alternatives are 
problem-adapted coordinate systems, e.g. \textsc{NRPy+} \cite{Ruchlin:2017com} 
or discontinuous Galerkin methods as in \textsc{SpECTRE} 
\cite{deppe_nils_2021_4734670,Kidder:2016hev} (see also 
\cite{Cao:2018vhw,Cai:2018cgd}). \change{}{Furthermore, it should be noted 
that some code frameworks, such as those based on the Carpet mesh refinement 
driver \cite{Schnetter:2003rb}, are technically capable of performing 
block-structured AMR. However, it can be cumbersome to use and these codes 
typically rely on moving-box type methods (e.g. the codes referenced in 
footnote \ref{fn:moving-box-codes}). }A brief overview of the history of NR 
codes  can be found in \cite{Sperhake:2014wpa}.},  which was first introduced 
in 2015  and uses the \chombo \cite{ChomboReport} library. While our methods 
apply directly to \grchombo, we believe many of the lessons we have learned 
are general and may be useful to researchers who work with other numerical 
relativity codes that make use of block-structured AMR, in particular those 
which rely on the Berger-Rigoutsos \cite{Berger:1991} style grid generation 
methods.

We demonstrate the utility of the techniques we have employed through a direct 
comparison of gravitational waveforms generated by binary black-hole inspiral 
and merger calculated by \grchombo and the more established \lean code which 
uses the aforementioned ``moving boxes'' style mesh refinement, and show that 
\grchombo is capable of achieving comparable production-level accuracy. We 
secondly apply AMR techniques to the evolution of several scalar field models 
which exhibit dynamics on a wide range of spatial and temporal scales, to 
demonstrate the relative advantages of several tagging criteria implemented in 
\grchombo.

This paper is organized as follows:
\begin{itemize}
    \item In \sref{sec:grchombo} we detail the computational framework of 
    \grchombo with a focus on the AMR aspects.
    \item In \sref{sec:amr-techniques} we discuss considerations for tagging 
    criteria in AMR grid generation.
    \item In sections \ref{sec:bh} and \ref{sec:osc} we illustrate how these 
    techniques are applied in practice to simulations of BH binaries and 
    spacetimes with a (self-interacting) scalar field. 
\end{itemize}

Our notation conventions are as follows. We use Greek letters 
$\mu,\nu,\ldots=0,1,2,3$ for spacetime indices and Latin letters 
$i,j,\ldots=1,2,3$ for spatial indices. We use a mostly plus signature 
$({-}{+}{+}{+})$ and geometric units $c=G=1$. When there is a potential for 
ambiguity between spacetime and purely spatial tensors (e.g. the Ricci scalar 
$R$), we prepend a ${}^{(4)}$ to denote the spacetime quantity. In the sections 
on black holes we set the mass scales with respect to the ADM mass of the 
spacetime, whereas for the section on scalar fields we set $\mu=mc/\hbar=1$, 
which then describes lengths relative to the scalar Compton 
wavelength\footnote{If we are interpreting the results in terms of a physical 
particle mass, this is equivalent to setting the value of $\hbar$ in the code. 
Note that in general $\hbar \neq 1$ in NR simulations (as this would imply that 
one unit in the length scale is equal to the Planck length $l_{\text{Pl}}$).}.

\section{Computational Framework}
\label{sec:grchombo}

This section provides a comprehensive update of the methodology discussed in 
section 2 of \cite{Clough:2015sqa}. See also \ref{sec:parallelization} for 
details on how \grchombo is parallelized.

\subsection{Mathematical equations and notational conventions}

\subsubsection{Evolution equations and gauge 
conditions}\label{sec:evolutionequations}

\grchombo implements the CCZ4 formulation \cite{Alic:2011gg,Alic:2013xsa} in 
order to evolve solutions of the Einstein equations, which we review below. We 
have found empirically with AMR that the inclusion of constraint damping terms 
in this formalism can be important to maintain accuracy; these mitigate the 
additional noise introduced by spurious reflections off the refinement 
boundaries due to the more complicated grid structures \cite{Baker:2005xe} 
(when compared with moving-box mesh refinement grids). The Z4 equation of 
motion with constraint damping and cosmological constant is 
\cite{Alic:2011gg}\footnote{Note the sign difference between the unit normal 
in \cite{Alic:2011gg} and here. We choose the sign which ensures the unit 
normal is future-directed.}
\begin{equation}\label{eq:Z4-covariant}
    {}^{(4)}R_{\mu\nu} - \Lambda g_{\mu\nu} + 2\nabla_{(\mu} Z_{\nu)} -
    2\kappa_1 n_{(\mu} Z_{\nu)}
    + \kappa_1(1 + \kappa_2)g_{\mu\nu}n_\alpha Z^\alpha 
    = 8\pi \left(T_{\mu\nu} - \frac{1}{2}g_{\mu\nu}T\right),
\end{equation}
where $\nabla$ is the Levi-Civita connection of the metric $g_{\mu\nu}$, 
${}^{(4)}R_{\mu\nu}$ is the Ricci tensor of $\nabla$, $\Lambda$ is the 
cosmological constant, $Z^\mu$ is the Z4 vector which vanishes on physical 
solutions of the Einstein equation, $n^\mu$ is the future-directed unit normal 
to the foliation of spatial slices, $\kappa_1$ and $\kappa_2$ are constant 
damping parameters, $T_{\mu\nu}$ is the energy-momentum tensor, and $T = 
g^{\alpha\beta}T_{\alpha\beta}$ is its trace.

In the standard $3+1$ decomposition of spacetime 
\cite{Arnowitt:1962hi,York1979}, the metric in adapted coordinates $(t,x^i)$ 
takes the form
\begin{equation}\label{eq:adaptedcoords}
    ds^2 = -\alpha^2\,\rmd t^2 + \gamma_{ij}(\rmd x^i + 
    \beta^i\,\rmd t)(\rmd x^j + \beta^j\,\rmd t), 
\end{equation}
where $\gamma_{ij}$ is the spatial metric, $\alpha=1/\Vert\rmd t\Vert$ is the 
\emph{lapse} function and $\beta^i$ is the \emph{shift} vector. The 
future-directed unit normal to the foliation is
\begin{equation}\label{eq:normal}
    n_\mu := -\alpha(\rmd t)_{\mu}, \qquad 
    n^\mu = \frac{1}{\alpha}\left(\partial_t^{\mu} - 
    \beta^k \partial_k^{\mu}\right),
\end{equation}
and the extrinsic curvature is
\begin{equation}
    K_{\mu\nu} := -\frac{1}{2}\left(\mathcal{L}_n \gamma\right)_{\mu\nu},
\end{equation}
which in adapted coordinates (\ref{eq:adaptedcoords}) becomes 
\begin{equation}
    K_{ij} = -\frac{1}{2\alpha} \left[
    \partial_t \gamma_{ij}-\beta^m\partial_m \gamma_{ij}
    -2\gamma_{m(i}\partial_{j)}\beta^m
    \right].
\end{equation}
In analogy to the spacetime metric, we decompose the energy-momentum tensor 
according to
\begin{equation}\label{eq:decomposed-EM-tensor}
    \rho := n_{\alpha} n_{\beta} T^{\alpha\beta},
    \quad
    S_i := -\gamma_{i\alpha}n_{\beta} T^{\alpha\beta},
    \quad
    S_{ij} := \gamma_{i\alpha}\gamma_{j\beta} T^{\alpha\beta},
    \quad
    S := \gamma^{ij}S_{ij}.
\end{equation}
We conformally rescale the spatial metric as\footnote{This is the main 
difference with the previous \grchombo paper \cite{Clough:2015sqa}, where 
$\chi^2$ was used in place of $\chi$.}
\begin{equation}
    \tilde{\gamma}_{ij} = \chi \gamma_{ij},\qquad 
    \tilde{\gamma}^{ij} = \frac{1}{\chi}\gamma^{ij},
\end{equation}
where $\chi = [\det (\gamma_{ij})]^{-1/3}$ so that the determinant of the 
conformal metric $\tilde{\gamma}_{ij}$ is unity. As in the BSSNOK formulation, 
we introduce the conformally rescaled traceless extrinsic curvature, 
\begin{equation}
    \tilde{A}_{ij} := \chi\left(K_{ij} - \frac{1}{3}K\gamma_{ij}\right),
    \label{eq:Aij-def}
\end{equation}
with $K=\gamma^{ij}K_{ij}$ and the \emph{conformal connection functions}
\begin{equation}\label{eq:conformal-connection-functions}
    \tilde{\Gamma}^i := \tilde{\gamma}^{jk}\tilde{\Gamma}^i_{jk} 
    = -\partial_j\tilde{\gamma}^{ij},
\end{equation} 
where $\tilde{\Gamma}^i_{jk}$ are the Christoffel symbols with respect 
to the conformal metric $\tilde{\gamma}_{ij}$. Finally, we decompose the Z4 
vector by defining\footnote{\label{fn:beforekappa1}The spatial projection is 
often denoted just $Z^i$ 
in the literature e.g. \cite{Alic:2011gg}, but we have changed the symbol in 
order to make clear that this is the \emph{projected} 3-vector as opposed to 
the spatial components of the Z4 vector} 
\begin{equation}\label{eq:Theta-Thetai-def}
    \Theta:=-n_\alpha Z^\alpha,\qquad \Theta^i := 
\gamma^{i}_{\phantom{i}\alpha}Z^\alpha.
\end{equation}
Rather than evolving the $\Theta^i$ directly, we instead incorporate them into 
a set of \emph{modified conformal connection functions}
\begin{equation}\label{eq:Gammahat-def}
    \hat{\Gamma}^i := \tilde{\Gamma}^i + 2\tilde{\gamma}^{ik}\Theta_k 
    = \tilde{\Gamma}^i + 2\frac{\Theta^i}{\chi}.
\end{equation}
Finally our complete set of dynamical variables are
\begin{equation}
\left\{\chi,\tilde{\gamma}_{ij},K,\tilde{A}_{ij},\Theta,\hat{\Gamma}^i\right\},
\end{equation}
and the CCZ4 evolution equations are\footnote{\label{fn:kappa1}Note that by 
default we make the modification $\alpha\kappa_1 \to \kappa_1$ mentioned in 
\cite{Alic:2013xsa} that allows stable evolution of BHs with $\kappa_3=1$. 
However, Eqs.~(\ref{eq:CCZ4-chi}-\ref{eq:CCZ4-Gamma}) do not have this 
modification.}

\begin{align}
    \partial_t\chi &= \beta^k\partial_k\chi + \frac{2}{3}\chi(\alpha K - 
    \partial_k\beta^k),\label{eq:CCZ4-chi}\\
    \partial_t\tilde{\gamma}_{ij} &= \beta^k\partial_k\tilde{\gamma}_{ij} 
    + \tilde{\gamma}_{ki}\partial_j\beta^k 
    + \tilde{\gamma}_{kj}\partial_i\beta^k -2\alpha \tilde{A}_{ij} 
    - \frac{2}{3}\tilde{\gamma}_{ij}\partial_k\beta^k,\\
    \begin{split}
        \partial_t K &= \beta^k\partial_k K + \alpha \left(\hat{R} 
        + K(K-2\Theta)\right) - 3\alpha\kappa_1(1+\kappa_2)\Theta 
        - \gamma^{kl}D_kD_l\alpha\\ 
        &\qquad
        + 4\pi\alpha (S - 3\rho) - 3\alpha\Lambda,
    \end{split}\\
    \begin{split}
            \partial_t\tilde{A}_{ij} &= \beta^k\partial_k \tilde{A}_{ij} 
        + \chi\left[-D_iD_j\alpha + \alpha(\hat{R}_{ij} 
        - 8\pi S_{ij}) \right]^{\text{TF}} \\
        &\qquad
        + \tilde{A}_{ij} \left[\alpha(K-2\Theta) 
        - \frac{2}{3}\partial_k\beta^k\right]
        + 2\tilde{A}_{k(i}\partial_{j)}\beta^k 
        - 2\alpha\tilde{\gamma}^{kl}\tilde{A}_{ik}\tilde{A}_{lj},
    \end{split}\\
    \begin{split}
        \partial_t\Theta &= \beta^k\partial_k\Theta 
        + \frac{1}{2}\alpha\left(\hat{R} - \tilde{A}_{kl}\tilde{A}^{kl} 
        + \frac{2}{3}K^2-2\Theta K\right)
        - \alpha\kappa_1\Theta(2 + \kappa_2) - \Theta^k\partial_k\alpha\\ 
        &\qquad
        - 8\pi \alpha \rho - \alpha\Lambda,
    \end{split}\\
    \begin{split}
        \partial_t\hat{\Gamma}^i &= \beta^k\partial_k\hat{\Gamma}^i 
        + \frac{2}{3}\left[\partial_k\beta^k\left(\tilde{\Gamma}^i 
        + 2\kappa_3\frac{\Theta^i}{\chi}\right)  
        - 2\alpha K\frac{\Theta^i}{\chi}\right]  
        - 2\alpha\kappa_1\frac{\Theta^i}{\chi}\\
        &\qquad
        + 2\tilde{\gamma}^{ik}(\alpha\partial_k\Theta 
        - \Theta\partial_k\alpha) - 2\tilde{A}^{ik}\partial_k\alpha 
        + 2\alpha\tilde{\Gamma}^i_{kl}\tilde{A}^{kl}\\
        &\qquad\qquad
        - \alpha\left[\frac{4}{3}\tilde{\gamma}^{ik}\partial_kK 
        + 3\tilde{A}^{ik}\frac{\partial_k\chi}{\chi}\right]
        - \left(\tilde{\Gamma}^k 
        + 2\kappa_3\frac{\Theta^k}{\chi}\right)\partial_k\beta^i\\
        &\qquad\qquad\qquad
        + \tilde{\gamma}^{kl}\partial_k\partial_l\beta^i 
        + \frac{1}{3}\tilde{\gamma}^{ik}\partial_l\partial_k\beta^l
        -16\pi \alpha\tilde{\gamma}^{ik}S_k,\label{eq:CCZ4-Gamma}
    \end{split}
\end{align}
where $D_i$ is the Levi-Civita connection on the spatial slice, 
$[\cdot]^{\text{TF}}$ denotes the trace-free part of the expression in square 
brackets, $\hat{R}_{ij}$ is the \emph{modified Ricci tensor}, given in terms of 
the normal Ricci tensor of $D_i$, $R_{ij}$ by
\begin{equation}\label{eq:modified-Ricci-def}
    \hat{R}_{ij} := R_{ij} + 2D_{(i}\Theta_{j)},
\end{equation}
and $\kappa_3$ is a third damping parameter. At each right-hand side 
evaluation, we construct the quantity $\Theta^i/\chi$ using the evolved 
$\hat{\Gamma}^i$ and $\tilde{\Gamma}^i$ computed from the metric and its 
derivatives in \eqref{eq:Gammahat-def}. The covariant second derivatives of 
the lapse are computed via 
\begin{align}
    \gamma^{kl}D_kD_l\alpha &= 
    \tilde{\gamma}^{kl}\chi\partial_k\partial_l\alpha 
    - \frac{1}{2} \tilde{\gamma}^{kl}\partial_k\alpha\partial_l\chi
    - \chi\tilde{\Gamma}^k\partial_k\alpha,\\
    \begin{split}
        D_iD_j\alpha &= \partial_i\partial_j\alpha 
        - \tilde{\Gamma}^k_{ij}\partial_k\alpha 
        + \frac{1}{2\chi}\left(\partial_i\alpha\partial_j\chi \right. \\
        &\qquad\left.+ \partial_i\chi\partial_j\alpha 
        - \tilde{\gamma}_{ij}\tilde{\gamma}^{kl}
        \partial_k\alpha\partial_l\chi\right).
    \end{split}
\end{align}
The modified Ricci tensor \eqref{eq:modified-Ricci-def} is given by\footnote{ 
Note the somewhat unconventional factor of $\chi^{-1}$ multiplying 
$R^\chi_{ij}$.}
\begin{equation}
    \hat{R}_{ij} = \tilde{R}_{ij} + \frac{1}{\chi}\left(R^{\chi}_{ij} 
    + R^Z_{ij}\right),
\end{equation}
with
\begin{align}
    \begin{split}
        \tilde{R}_{ij} &= 
        -\frac{1}{2}\tilde{\gamma}^{kl}\partial_k\partial_l\tilde{\gamma}_{ij} 
        + \tilde{\gamma}_{k(i}\partial_{j)}\hat{\Gamma}^k
        +\frac{1}{2}\hat{\Gamma}^k\partial_k \tilde{\gamma}_{ij}\\
        &\qquad 
        +\tilde{\gamma}^{lm}\left(\tilde{\Gamma}^k_{li}\tilde{\Gamma}_{jkm} 
        + \tilde{\Gamma}^k_{lj}\tilde{\Gamma}_{ikm} 
        + \tilde{\Gamma}^k_{im}\tilde{\Gamma}_{klj}\right), 
    \end{split}
    \\
    \begin{split}
        R^\chi_{ij} &= \frac{1}{2}\left[\tilde{D}_i\tilde{D}_j\chi 
        +\tilde{\gamma}_{ij}\tilde{\gamma}^{kl}
        \tilde{D}_k\tilde{D}_l\chi\right] \\
        &\qquad -\frac{1}{4\chi}\left[\partial_i\chi\partial_j\chi 
        + 3\tilde{\gamma}_{ij}\tilde{\gamma}^{kl}
        \partial_k\chi\partial_l\chi\right],
    \end{split}
    \\
        R^Z_{ij} &= 
        \frac{\Theta^k}{\chi}\left(\tilde{\gamma}_{ik}\partial_j\chi 
        + \tilde{\gamma}_{jk}\partial_i\chi 
        - \tilde{\gamma}_{ij}\partial_k\chi
        \right),
\end{align}
and $\tilde{\Gamma}_{ijk}=\tilde{\gamma}_{il}\tilde{\Gamma}^l_{jk}$.
Finally, the \emph{modified Ricci scalar} is given by the trace of the modified
Ricci tensor:
\begin{equation}
    \hat{R} = \chi \tilde{\gamma}^{kl}\hat{R}_{kl}.
\end{equation}
By default, \grchombo uses the damping parameters$^{\ref{fn:kappa1}}$
\begin{equation}
    \alpha\kappa_1=0.1,\;\kappa_2=0\text{ and }\kappa_3=1.
    \label{eq:default-ccz4-damping}
\end{equation}

To close the system (\ref{eq:CCZ4-chi})-(\ref{eq:CCZ4-Gamma}), we need to 
specify gauge conditions for the lapse and shift. Although the structure of the 
code allows easy modification of the gauge, by default we use a 
\emph{Bona-Masso}-type slicing condition \cite{Bona:1994dr} of the form
\begin{equation}
    \partial_t\alpha = a_1\beta^k\partial_k\alpha 
    - a_2\alpha^{a_3}(K-2\Theta),
    \label{eq:Bona-Masso-slicing}
\end{equation}
where $a_1$, $a_2$ and $a_3$ are constant parameters. Note that this reduces to 
the familiar 1+log slicing in the case
\begin{equation}
    a_1 = 1,\;a_2 = 2\text{ and }a_3 = 1,
    \label{eq:default-lapse-params}
\end{equation}
which is the default. For the shift, we use the \emph{Gamma-driver} shift 
condition \cite{Campanelli:2005dd, Baker:2005vv} in the form:
\begin{align}
    \partial_t\beta^i &= b_1\beta^k\partial_k\beta^i + b_2 B^i, 
    \label{eq:Gamma-driver-1}\\
    \partial_tB^i &= b_1(\beta^k\partial_kB^i 
    - \beta^k\partial_k\hat{\Gamma}^i)  
    + \partial_t\hat{\Gamma}^i - \eta B^i,
    \label{eq:Gamma-driver-2}
\end{align}
where $b_1$, $b_2$ and $\eta$ are constant parameters. By default we take $b_1 
= 0$ and $b_2 = 3/4$, whereas the value of $\eta$ depends on the specific 
configuration considered (typically $\mathcal{O}({1/M_{\text{ADM}}})$). 
Together, these gauge conditions are commonly referred to as the \emph{moving 
puncture gauge}.

We enforce the tracelessness of $\tilde{A}_{ij}$ \eqref{eq:Aij-def} before 
every RHS evaluation and at the end of each full timestep. Furthermore, we also 
enforce the condition $\chi \geq \chi_{\min}$ and $\alpha \geq \alpha_{\min}$ 
before every RHS evaluation and at the end of each full timestep, where the 
default values are $\chi_{\min} = 10^{-4} = \alpha_{\min}$, in order to ensure 
these variables do not become arbitrarily small or negative due to 
discretization error\footnote{Typically any values that are affected by this 
procedure lie behind a horizon so are causally disconnected from most of the 
computational domain. In cosmological simulations the conformal factor is 
directly related to the scale factor and thus in rapidly expanding or 
contracting spacetimes this limit may need to be adjusted.}.

As is common when using the moving puncture gauge with BH spacetimes, \grchombo 
can track the position of the puncture(s) $\mathbf{x}_p$ by solving 
\cite{Alcubierre:2008co}
\begin{equation}
    \frac{\rmd x^i_p}{\rmd t} = -\beta^i(\mathbf{x}_p),
    \label{eq:puncture-tracking}
\end{equation}
which is integrated using the trapezium rule at user-specified time intervals.

During evolutions, we monitor the violations of the Hamiltonian and momentum 
constraints which are given by
\begin{align}
    \mathcal{H} &:= R + \frac{2}{3}K^2 
    - \tilde{A}_{kl}\tilde{A}^{kl}-2\Lambda-16\pi\rho,
    \label{eq:Ham-constraint}\\
        \mathcal{M}_i &:= \tilde{\gamma}^{kl}\left(\partial_k \tilde{A}_{li} 
        -2\tilde{\Gamma}^m_{l(i}\tilde{A}_{k)m}
        -3\tilde{A}_{ik}\frac{\partial_l\chi}{2\chi}\right)
        -\frac{2}{3}\partial_iK - 8\pi S_i.
    \label{eq:mom-constraint}
\end{align}
In the continuum limit these two quantities should vanish. From the Bianchi 
identities, if the constraints vanish on the initial data surface, they vanish 
throughout the spacetime \cite{Frittelli:1996nj}. However, given errors due to our 
discretization of the equations of motion, we expect there to be small 
violations of the constraints in our numerical simulations. We often monitor 
the constraint violation to assess the accuracy of a given run. 

In the case of non-vacuum spacetimes, one can normalize the constraints with 
some measure (e.g. the maximum) of their matter sources, i.e.~$\mathcal{H}$ with 
$\rho$ and $\mathcal{M}_i$ with $S_i$.

\subsubsection{Initial Data}
\label{sec:initial-data}

For BH binaries, we have integrated the \textsc{TwoPunctures} spectral solver 
\cite{Ansorg:2004ds} into \grchombo. This provides binary puncture data 
\cite{Brandt:1997tf} of Bowen-York type \cite{Bowen:1980yu}; the version we use 
also incorporates the improvements described in \cite{Paschalidis:2013oya} that 
allows for fast spectral interpolation of the pseudospectral solution onto, 
e.g. a Cartesian grid. Besides \textsc{TwoPunctures}, \grchombo includes a 
class for non-spinning binary Bowen-York data with an approximate solution of 
the Hamiltonian constraint for the conformal factor \cite{Dennison:2006nq}, 
which is valid in the limit of small boosts $|\mathbf{P}|\ll M$, where 
$\mathbf{P}$ is the initial momentum of an individual black hole and $M$ is some 
measure of the total mass.

We also provide initial data for Kerr BHs using the formulation in 
\cite{Liu:2009al}, which allows for the evolution of near-extremal BHs 
within the moving puncture approach to black hole evolution.

For matter spacetimes, one must in general solve the Hamiltonian and momentum 
constraints numerically to obtain valid initial conditions for the metric that 
correspond to the energy and momentum distributions which are chosen (including 
BHs using the methods above, where required). For time-symmetric spherically 
symmetric initial data this can be done using shooting methods, as for the 
axion star data used in \sref{sec:osc} and previous works 
\cite{Nazari:2020fmk,Muia:2019coe,Widdicombe:2019woy,Widdicombe:2018oeo, 
Clough:2018exo,Dietrich:2018bvi,Helfer:2018vtq,Helfer:2016ljl}. In 
inhomogeneous cosmological spacetimes, numerical solutions for an initial 
matter configuration should be obtained using relaxation or multigrid methods, 
as in 
\cite{Joana:2020rxm,Aurrekoetxea:2019fhr,Clough:2017efm,Clough:2016ymm}%
\footnote{Most of these works assume an initial time symmetry in order to avoid 
solving the momentum constraints. A fully general multigrid initial condition 
solver for \grchombo that solves the coupled constraint system for any scalar 
field distribution with and without BHs is currently under development.}.

\subsection{Code features}

\subsubsection{Discretization and time-stepping}
\label{sec:discretization}

\grchombo evolves the CCZ4 equations using the method of lines with a standard 
fourth-order Runge-Kutta method (RK4). For the spatial discretization, we 
typically use either fourth or sixth order centered stencils except for the 
advection terms, for which we switch to lopsided stencils of the same order 
depending on the sign of the shift vector component. For completeness, the 
expressions for the stencils are provided in \ref{sec:stencils}.

Finite difference methods can often introduce spurious high-frequency modes, 
particularly when using adaptive mesh refinement and regridding. To ameliorate 
this, \grchombo uses $N=3$ \emph{Kreiss-Oliger} (KO) dissipation 
\cite{Kreiss:1973}; at every evaluation of the right-hand side (RHS) for an 
evolution variable $F$, we add the term
\begin{equation}
    \frac{\sigma}{64\Delta x}\left(F_{i-3} - 6F_{i-2} + 15F_{i-1} - 20F_i
    +15F_{i+1}-6F_{i+2} + F_{i+3}\right)
\end{equation}
to the RHS, where $\Delta x$ is the relevant grid spacing (cf. 
\eqref{eq:grid-spacing}). A von Neumann stability analysis 
\cite{Alcubierre:2008co} shows that this scheme, when applied to the trivial 
partial differential equation (PDE) $\partial_t F=0$, is linearly stable only if
\begin{equation}
    0\leq \sigma \leq \frac{2}{\alpha_C}, \label{eq:CFL}
\end{equation}
where $\alpha_C=\Delta t/\Delta x$ is the \emph{Courant-Friedrichs-Lewy} 
factor, which we typically set to $1/4$, and $\Delta t$ is the size of the 
timestep\footnote{Note that the stability analysis makes a number of 
assumptions and problems can begin to appear towards the upper end of the range 
(\ref{eq:CFL}). We have observed that a typical symptom of an instability due 
to too large $\sigma$ is the appearance of a checkerboard-like pattern in 
otherwise spatially homogeneous regions of the spacetime.}. Note that we always 
use $N=3$ KO dissipation, independent of the order of spatial discretization. 
Naively, one might question this choice for sixth order spatial derivative 
stencils as the conventional wisdom is to pick $N$ such that $2N-1>m$, where 
$m$ is the order of the finite difference scheme (see, for example, 
\cite{Alcubierre:2008co}). However, in this case, what matters is the 
order of the time stepping which is still fourth order\footnote{Also note that 
Theorem 9.1 in \cite{Kreiss:1973} only refers to the order of the time 
stepping.}, hence the dissipation operator does not ``spoil'' the convergence 
properties of the scheme. This is consistent with the approach discussed in 
section 3.2 of \cite{Husa:2007hp}.

\subsubsection{Berger-Rigoutsos AMR}
\label{sec:BR-AMR}
\begin{figure}[t]
    \centering
    
    \includegraphics[width=0.7\columnwidth]{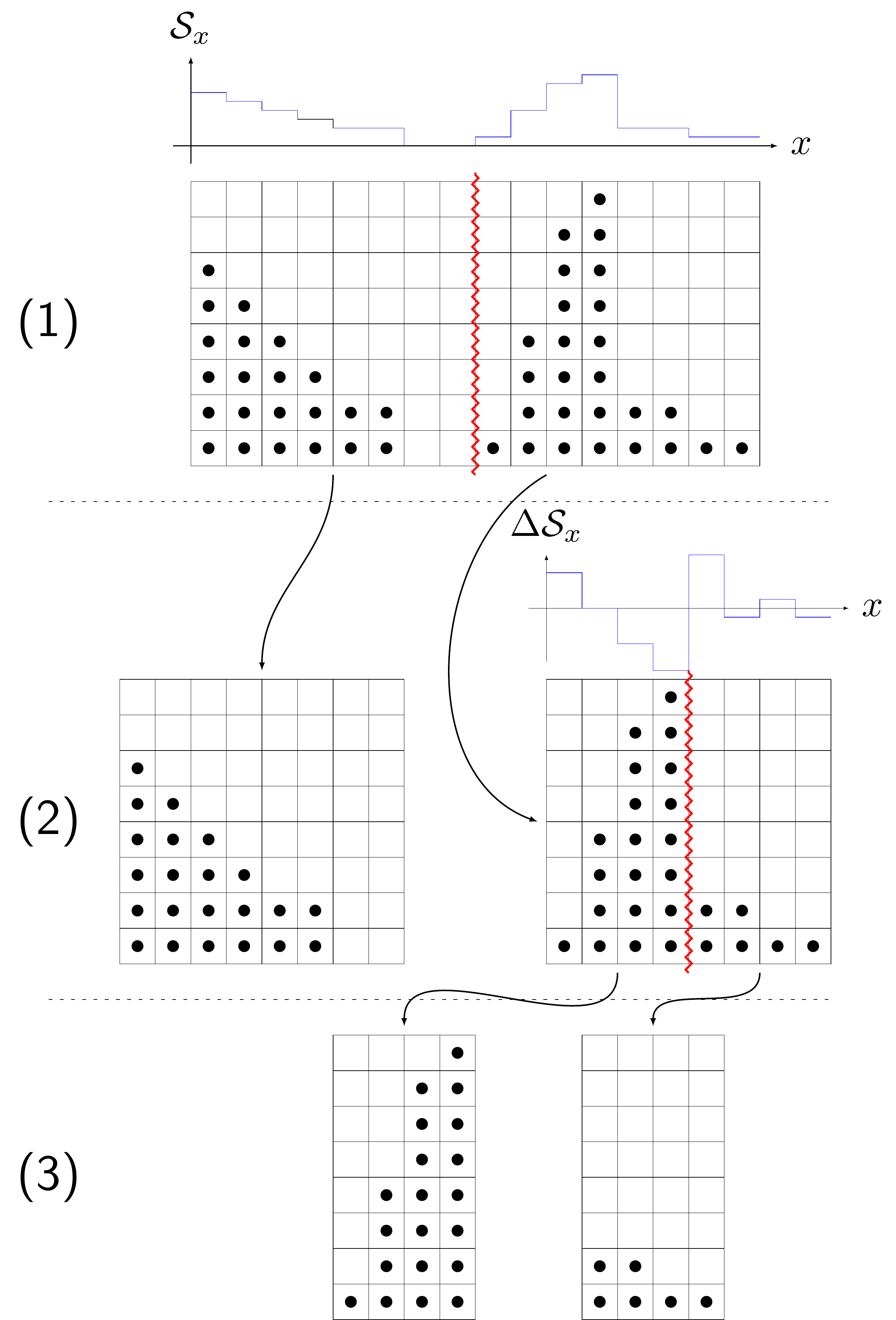}
    \caption{
    Schematic illustration of the partitioning algorithm. For simplicity, we 
show
    a 2D grid and only consider partitioning in the $x$ direction. 
    The cells tagged
    for refinement are indicated with $\bullet$.
    In (1), the 
    signature $\mathcal{S}_x$ is computed and two ``holes'' are found where 
    the signature vanishes.
    The line (plane in 3D) of partition is then at the hole with the 
    largest index (rightmost). The result of the partitioning is shown in (2).
    To partition the right box in (2), the signature is computed but this time
    there are no holes so the algorithm looks for zero crossings of the 
    discrete Laplacian of the signature $\Delta\mathcal{S}_x$. The zero-crossing
    with largest change is then selected.
    This algorithm terminates once all boxes
    have reached the required fill ratio $\epsilon_{\rm FR}$.}
    \label{fig:berger-rigoutsos-partitioning}
\end{figure}
In \grchombo, the grid comprises a hierarchy of cell-centered Cartesian meshes 
consisting of up to $\lmax+1$ refinement levels labeled\footnote{Note that the 
finest level that exists may be less than $\lmax$.} $l=0,\ldots,\lmax$ each 
with grid spacing
\begin{equation}
    \Delta x_l=2^{\lmax-l}\Delta x_{\lmax}=\Delta x_0/2^l.
    \label{eq:grid-spacing}
\end{equation}
\grchombo uses  block-structured AMR, so each refinement level is split into 
boxes which are distributed between the CPUs as described in 
\ref{sec:parallelization}.

At regridding or initial grid creation, on a given refinement level $l$, cells 
are tagged for refinement according to a \emph{tagging criterion} 
$C=C(\mathbf{i})$. In a given cell with indices $\mathbf{i}=(i,j,k)$ and 
corresponding Cartesian coordinates\footnote{ Note that the indices here are of 
the discrete cells on the grid as opposed to spacetime components.}
$\mathbf{x}=(x_i,y_j,z_k)$, if $C(\mathbf{i}) > \tau_R$, where $\tau_R$ is a 
pre-specified threshold value (which may vary with $l$), then the cell is 
tagged for refinement. We discuss techniques to design a suitable tagging 
criterion and aspects to consider in \sref{sec:amr-techniques}.

In block-structured AMR, the main challenge after tagging cells is finding an 
efficient algorithm to \emph{partition} the cells that require refinement into 
blocks or boxes. \grchombo uses \chombo's implementation of the 
Berger-Rigoutsos grid generation algorithm \cite{Berger:1991} in order to do 
this. 

For this purpose, we define the \emph{block factor} as the number of cells that 
must divide the side lengths of all blocks; it is a specifiable parameter. 
Furthermore these side lengths must not exceed the specified maximum box size. 
In order to enforce the block factor on level $l+1$, starting with the tagged 
cells on level $l$, \chombo generates a temporary new set of tagged cells on a 
virtual coarser level $l-n$ where $n$ is chosen such that the length of one 
cell on level $l-n$ corresponds to the block factor on level $l+1$\footnote{For 
example if the block factor is 4, then $n=1$ since the refinement ratio is 2 
and $2^{(l+1)-(l-1)}=4$. Note that this means the block factor must be a power 
of 2.}. The new set of coarser tags are derived using a global OR operation, 
i.e. as long as any of the $l$ level cells corresponding to the coarser level 
cell is tagged, the virtual coarser level cell will be tagged. \chombo then 
applies the Berger-Rigoutsos partitioning algorithm on this coarser level to 
construct boxes of grids which obey both the desired block factor and maximum 
box size. We typically choose both to be a multiple of the processor vector 
width for optimal performance.

For completeness, we next review the Berger-Rigoutsos algorithm (see also 
figure \ref{fig:berger-rigoutsos-partitioning}). We find the minimum box that 
encloses all of the tagged cells on this level. Let $T(\mathbf{i})$ be the 
tagging indicator function defined by
\begin{equation}
    T(\mathbf{i}) =
    \begin{cases}
    1, & \text{if } C(\mathbf{i}) > \tau_R,\\
    0, & \text{otherwise}.
    \end{cases}
    \label{eq:tagging-switch}
\end{equation}
and define the \emph{signatures} or {traces} of the tagging by
\begin{align}
    \mathcal{S}_x(i) &:= \sum_{j,k} T(\mathbf{i}) 
        = \int T(\mathbf{i})\,\mathrm{d}y\,\mathrm{d}z,\\
    \mathcal{S}_y(j) &:= \sum_{i,k} T(\mathbf{i}) 
        = \int T(\mathbf{i})\,\mathrm{d}x\,\mathrm{d}z,\\
    \mathcal{S}_z(k) &:= \sum_{i,j} T(\mathbf{i})
        = \int T(\mathbf{i})\,\mathrm{d}x\,\mathrm{d}y.
\end{align}
First, we look for ``holes'' in the signatures i.e.~if there exist $i$, $j$ or 
$k$ for which $\mathcal{S}_x(i)$, $\mathcal{S}_y(j)$ or $\mathcal{S}_z(k)$ 
vanish which corresponds to there being no tagged cells along the plane 
orthogonal to the signature direction. If there are holes, we choose the one 
with largest index over all the dimensions (since it is more efficient to have 
fewer big boxes than more small boxes) as the plane of partition. If there are 
no holes, we next look for inflections (see \cite{Berger:1991} and their 
discussion of figure 10 for details) in the signatures by computing their 
discrete Laplacian, for example,
\begin{equation}
    \Delta\mathcal{S}_x(i) 
        = \mathcal{S}_x(i-1) -2\mathcal{S}_x(i)+\mathcal{S}_x(i+1),
\end{equation}
and searching for zero-crossings in $\Delta\mathcal{S}_x(i)$. Heuristically, 
this corresponds to a rough boundary between tagged and untagged regions; cf. 
the partitioning in step (2) of figure 
\ref{fig:berger-rigoutsos-partitioning}). If there are inflections, then, in 
each direction, we pick the inflection with the greatest difference, for 
example, 
\begin{equation}
    |\delta(\Delta\mathcal{S}_x(i))| 
        = |\Delta\mathcal{S}_x(i-1)-\Delta\mathcal{S}_x(i)|,
\end{equation}
As for the holes, we then pick the greatest inflection index over all the 
dimensions as our plane of partition. If there are no holes or inflections in 
the signatures, we simply split the box along the midpoint of the direction 
with the longest side.

After partitioning, we check whether the partition is sufficiently 
\emph{efficient}, specifically whether the proportion of tagged cells to all 
cells in the partition exceeds a user-specified \emph{fill ratio} threshold, 
$\epsilon_{\text{FR}}<1$ and that the lengths of the boxes are at most the 
pre-specified maximum box size (which we choose in order to ensure sufficient 
load balancing). If these tests are passed then we accept the partition and, if 
not, we continue to partition recursively discarding any boxes that do not 
contain tagged cells.

Note that, whilst a higher value of $\epsilon_{\text{FR}}$ will result in a 
more efficient partition in the sense that there will be a greater ratio of 
tagged to untagged cells, this is not always the most computationally efficient 
choice as there are greater overheads with smaller boxes (for example, there 
will be more ghost cells). There could also be more fluctuation in the 
structure of the grids between consecutive regrids which may result in greater 
noise. Although the optimal fill ratio depends on the particular physical 
problem and the computational resources, we typically use 
$\epsilon_{\text{FR}}=0.7$.

Finally the boxes in the partition are refined, that is, they are defined on 
the next finer level ($l+1$) with twice the resolution \eqref{eq:grid-spacing}. 
For newly created regions on this finer level, we interpolate the data from the 
coarser level using fourth-order interpolation.

The regridding process starts on the finest level that currently exist (or at 
most level $\lmax-1$) and works up the hierarchy on increasingly coarse levels 
until the base level, from whose timestep the regrid was called, is reached 
(which need not be $l=0$). It is therefore only possible to add a single extra 
level (up to $\lmax$) at each regrid. After the regrid on level $l$, the union 
of the set of cells in the new boxes on this level (plus an additional 
pre-specified \emph{buffer region}) with the set of cells tagged on level $l-1$ 
is used as the final set of tagged cells on level $l-1$ in order to ensure 
\emph{proper nesting}\footnote{By proper nesting we mean that
\begin{enumerate}[label=(\roman*)]
    \item The physical region corresponding to a level $l-1$ cell must be fully 
    refined or not refined at all, that is it must be completely
    covered by level $l$ cells and not partially refined.
    \item There must be at least one level $l$ cell between the boundary of 
    $l+1$ and the boundary of level $l$ except at the boundary of the entire 
    computational domain. In practice we even need two such buffer cells 
    (corresponding to 4 cells on level $l+1$) for fourth and sixth-order spatial 
    stencils.
\end{enumerate}
}
\cite{Berger:1989}.
This also ensures that cells on coarser levels will be tagged if any of their 
corresponding finer level cells are tagged.

The frequency of regridding is user-specifiable on each refinement level 
$l<\lmax$, though, since a regrid on level $l=l^{\prime}$ enforces a regrid on 
all levels $l^\prime \leq l < \lmax$, for problems without rapidly varying (in 
time) length scales, it is usually sufficient to regrid every few timesteps on 
one of the more coarse levels (e.g. for compact object binaries). Not only does 
reducing the frequency of regridding reduce the computational cost, but since 
regridding introduces errors/noise due to interpolation, we have also found 
that this can improve the accuracy of the simulation.

The Courant condition limits the size of the maximum time steps one can take on 
the finer levels. Rather than evolving all refinement levels with the same 
timestep, we use \emph{subcycling} by following the Berger-Colella evolution 
algorithm \cite{Berger:1989}, which we now review. As the algorithm is 
recursive, we can consider evolving a set of coarser and finer grids at level 
$l$ and $l+1$ respectively in the AMR grid hierarchy. First, one time step is 
taken on the coarser grids (i.e. those at level $l$). One then evolves the 
finer (level $l+1$) grids for as many time steps until they have advanced to 
the same time as the coarse grid. As we have hard-coded the time steps on each 
level, $\Delta t_l$, so that $\Delta t_l = \Delta t_{l-1} / 2$, the grids on 
level $l+1$ will then take two time steps after the grids on level $l$ take one 
time step. After level $l+1$ has ``caught up'' with level $l$, the mean of the 
data in the [$2^3=8$] cells covering a single level $l$ cell is calculated and 
this value is copied back onto level $l$.\change{}{ Note that this particular
procedure is only second-order accurate in contrast to the restriction 
operation in a vertex-centred code which requires no approximation. This
may partially explain some of the difference in convergence orders we observed
between \grchombo and the vertex-centred code \lean in section \ref{sec:bh}.}

The ghost cells at the interface between the finer and coarser grids are set by 
interpolating the values of the coarser grid in both space (due to the cell 
centered grid) and time (due to the requirement for intermediate values in the 
RK4 timestepping). The time interpolation is achieved by fitting the 
coefficients of a 3rd order polynomial in $t$ using the values obtained at each 
substep of the RK4 timestepping on the relevant cells of the coarser level (see 
\cite{Colella:2011} for more detail).\footnote{ An alternative approach 
would be the use of larger ghost zones in the finer level, with the outer ones 
discarded at each RK step (for example, see section 2.3 of 
\cite{Schnetter:2003rb}). One disadvantage here is the extra memory use, 
especially beyond the fixed-box-hierarchy case. }

\subsubsection{Boundary Conditions}
\label{sec:boundary-conditions}

\grchombo implements several classes of boundary conditions, including:
\begin{itemize}
    \item Periodic - evolution variables $\varphi$ obey $\varphi(x+L) = 
    \varphi(x)$ in some or all Cartesian directions.
    \item Static - boundary values are fixed at their initial values.
    \item Reflective - one uses the symmetry of the spacetime to reduce the 
    volume evolved. For example, in a simple equal mass head-on BH merger, one 
    needs only 1/8 of the domain; the rest can be inferred from the evolved 
    values \cite{Bondarescu:2006sk}. Note that each  evolution variable has a 
    different parity across each reflective boundary.
    \item Extrapolating - both zeroth and first order schemes (by radial 
    distance from a central point). These are especially useful for variables 
    which asymptote to a spatially uniform but time varying value (see 
    \cite{Bamber:2020bpu, Traykova:2021dua, Clough:2021qlv}).
    \item Sommerfeld/radiative - these permit (massless) outgoing radiation to 
    leave the grid without reflections by 
    assuming a solution of the form $\varphi = \varphi_0 + u(r-t)/r$ at the 
    boundaries, where $u$ can be any arbitrary function and $\varphi_0$ is a 
    constant asymptotic value (see section 5.9 of \cite{Alcubierre:2008co}).
\end{itemize}

\subsubsection{Interpolation and wave extraction}
\label{sec:amrinterp-wave}

\grchombo features an AMR interpolator which allows the user to interpolate any 
grid variable, its first derivatives and any second derivative at an arbitrary 
point within the computational domain. The AMR interpolator starts searching for 
the requested points on the finest available level and then progresses down the 
hierarchy to increasingly coarse levels until all requested points are found in 
order to ensure the most accurate result. It supports interpolation via an 
arbitrary interpolation algorithm. The provided algorithms include Lagrange 
polynomial interpolation up to arbitrary order (although this is limited by the 
number of available ghost cells), using the algorithm of 
\cite{Fornberg:1998} to calculate the stencils on the fly and then 
memoizing 
them in order to increase efficiency, and ``nearest neighbour'' interpolation. 
By default, and in the remainder of this paper, we use fourth order Lagrange 
polynomial interpolation.

\grchombo includes tools for the extraction of data over an arbitrary 
user-defined 2D surface, built on top of the AMR interpolator, with spherical 
and cylindrical surfaces being implemented as examples. There is also a 
user-friendly interface for integrating arbitrary functions of the extracted 
grid variables and their derivatives over the surface using either the trapezium 
rule, Simpson's rule, Boole's rule or the midpoint rule in each surface 
coordinate direction.

Often the most important outputs of a numerical relativity simulation are the 
calculated gravitational waves. To that end, \grchombo uses the Newman-Penrose 
formalism \cite{Newman:1961qr}. We describe the formulae for calculating the 
relevant Weyl scalar $\Psi_4$ (including terms arising from the Z4 vector) in 
\ref{sec:psi4-appendix}. We use the extraction routines described 
above to interpolate the values of $\Psi_4$ on multiple spheres of fixed 
coordinate radius and then determine the modes $\psi_{lm}$ with respect to the 
spin-weight $-2$ spherical harmonics ${}_{-2}Y^{lm}$ using the formula
\begin{equation}
    r_{\text{ex}}\psi_{lm} 
    = 
    \oint_{S^2}r_{\text{ex}}\Psi_4|_{r=r_{\text{ex}}}
    \left[{}_{-2}\bar{Y}^{lm}\right]\,\rmd \Omega,
\end{equation}
where $\rmd \Omega = \sin\theta\,\rmd\theta\,\rmd\phi$ is the area element on 
the unit sphere $S^2$. We use the trapezium rule for the integration over 
$\phi$ (since the periodicity means that any quadrature converges exponentially 
\cite{Poisson:1827}) and Simpson's rule for the integration over $\theta$.


\section{Considerations for tagging criteria used for grid generation}
\label{sec:amr-techniques}

We have found that the choice of tagging criteria can greatly impact the 
stability and accuracy of a given simulation. Here we mention several factors 
to consider when designing tagging criteria for use in \grchombo and other 
codes with similar AMR algorithms. We also provide some explicit examples of 
tagging criteria and discuss their relative merits.

\subsection{Buffer Regions}
\label{sec:buffer-regions}

One of the problems of many tagging criteria we have tried is that they can 
often introduce several refinement levels over a relatively small distance in 
space. This leads to the boundaries of these refinement levels being 
particularly close to one another. Due to the errors introduced by 
interpolation at these boundaries, they can add spurious reflections or noise. 
This is exacerbated when other refinement boundaries are nearby, allowing for 
this noise to be repeatedly reflected and even amplified before it has time to 
dissipate (e.g.~via Kreiss-Oliger dissipation -- see 
\sref{sec:discretization}). A particularly simple way to mitigate this 
problem is to increase the \emph{buffer regions}, i.e.~the number of cells 
$n_{\rm B}$ between refinement levels. Since the regridding algorithm starts at 
the finest level and works up the hierarchy to coarser levels (see 
\sref{sec:BR-AMR}), increasing this parameter actually increases the size 
of the coarser levels rather than shrinking the finer levels in order to 
enforce this buffer region restriction.

\subsection{Considerations for black-hole spacetimes}
\label{sec:black-hole-tagging}

Here we describe several techniques that we have used when creating tagging 
criteria to evolve black-hole spacetimes. The major complication with evolving 
black holes is that they have an event horizon. In practice, it is often 
challenging to find the true event horizon, which would require tracing 
geodesics through the full evolution of the spacetime. Therefore NR simulations 
typically consider the location of the apparent horizon instead, which always 
lies inside the event horizon \cite{hawking1973large}. Mathematically, the 
region within an apparent horizon is causally disconnected from its exterior. 
For a given numerical approximation, however, artifacts from the discretization 
can propagate from behind the horizon and contaminate the rest of the 
computational domain.

As a consequence of this superluminal propagation of numerical noise, we often 
find that \grchombo simulations of BHs are particularly sensitive to the 
presence of refinement boundaries. One should avoid adding refinement within 
the horizon (which in any case is unobservable and not usually of interest), 
but problems are particularly severe where a refinement boundary intersects the 
apparent horizon. In such cases we have observed significant phase inaccuracies 
and drifts in the horizon area (some even violating the second law of black 
hole mechanics). Similar problems may occur if a refinement boundary is close 
to but does not intersect the horizon. In order to avoid these issues, we 
typically enforce the tagging of all cells within the horizon plus a buffer 
radius up to a maximum level $l_{\text{BH}}^{\max}$ (which need not necessarily 
be $\lmax$ and may differ for each BH in the simulation). If $r_p$ is the 
coordinate distance from the puncture of a BH of mass $M_{\text{BH}}$ in a 
spacetime with total mass $M\sim 2M_{\text{BH}}$, then, for $\eta\sim1/M$ in 
the moving puncture gauge 
(\ref{eq:Bona-Masso-slicing}-\ref{eq:Gamma-driver-2}), after the initial gauge 
adjustment the apparent horizon is at approximately $r_p = M_{\text{BH}}$ (see 
figure 4 in \cite{Bruegmann:2006ulg}). Guided by this approximation, we can tag 
all cells with $r_p<M_{\text{BH}}+b$, where $b$ is a pre-specified parameter. 
Although one might think choosing $b\propto M_{\text{BH}}$ for each BH might be 
the most sensible choice for unequal mass configurations, we have found larger 
BHs less sensitive to the presence of refinement boundaries. Thus, choosing 
$b\propto M$ the same for each BH in a binary usually works sufficiently well.

Increasing the size of the buffer regions between refinement boundaries by 
adjusting $n_{\text{B}}$ as discussed in \sref{sec:buffer-regions} can help 
to keep refinement boundaries sufficiently spaced apart. However, we have also 
separately enforced the spacing out of refinement boundaries by doubling the 
radius of the second and third finest levels covering a BH. This leads to 
tagging cells on level $l$ (to be refined on level $l+1$) with
\begin{equation}
    r_p<(M_{\text{BH}}+b)2^{\min(l^{\max}_{\text{BH}}-l-1,2)}
    \label{eq:spherical-puncture-tagging}
\end{equation}

In spacetimes where BH horizons are dynamical (often the target of AMR 
simulations), one can in principle use  the locations of apparent horizons to 
define tagged regions. However, rather than incorporating the output of a 
horizon finder into the tagging criterion, a  simpler and in most cases equally 
effective method can be obtained from using contours of the conformal factor 
$\chi$ and tagging regions with $\chi<\chi_0$, where $\chi_0$ is a prespecified 
threshold value which may vary on each refinement level. This gives a 
reasonably robust and general method of identifying the approximate locations 
of horizons. Further details on precise values and their dependence on the BH 
spin are given in \ref{sec:chihorizon}.

\subsection{Asymmetric grids}
\label{sec:asymmetric-grids-tagging}

The grid-generation algorithm (\sref{sec:BR-AMR}) is inherently asymmetric, 
for example, it picks the ``hole'' with \emph{largest} index as the partition 
plane. This means that even if the tagging has symmetries, the grids themselves 
may not obey the same or any symmetries. For example, whilst one might expect 
that, for tagging cells with \eqref{eq:spherical-puncture-tagging}, the 
grids would have reflective symmetry in all three coordinate directions, this 
is often not the case, particularly for larger $\epsilon_{\text{FR}}$. This 
asymmetry can lead to undesirable behaviour. For example, when simulating the 
head-on collision of two BHs with no symmetry assumptions (as described in 
\sref{sec:boundary-conditions}) with the tagging of 
\eqref{eq:spherical-puncture-tagging}, the punctures can deviate slightly 
from the collision axis. We can ``fix'' this asymmetry by replacing 
\begin{equation}
    r_p\to \varrho=\max(|x-x_p|,|y-y_p|,|z-z_p|)
    \label{eq:box-tagging}
\end{equation}
in  \eqref{eq:spherical-puncture-tagging} so that the tagged regions are 
boxes rather than spheres (this tagging is similar to what is done in some 
moving-box style mesh refinement codes). Whilst there is inevitably a loss of 
efficiency from this choice, this is typically outweighed by the reductions in 
error achieved. Clearly, this approach pushes the AMR method in the direction 
of a moving boxes approach; in practice, we therefore apply it predominantly to 
BH simulations but not for more complex matter structures that require the full 
flexibility of AMR.

\subsection{Using truncation error for tagging cells}
\label{sec:truncation-tagging}

Truncation error tagging was introduced by Berger \textit{et 
al.}~\cite{Berger:1984zza}. We have implemented truncation error tagging in 
\grchombo by using a \emph{shadow hierarchy} (e.g. \cite{pretorius_phd}). In 
this scheme, we estimate the truncation error on a grid at level $l$ by 
comparing the solution of a specially chosen variable $f$ on that level to the 
coarser level directly ``beneath'' it on the grid:
\begin{equation}
\label{eq:trunc_error}
    \tau_{l,f}(\mathbf{i})
    =
    \left|f_{l}(\mathbf{i}) - f_{l-1}(\mathbf{i})\right|
    .
\end{equation}
We note that the error \eqref{eq:trunc_error} clearly must be computed before 
we average the finer grid values onto the coarser grid. As \chombo uses a 
cell-centered scheme, in order to compare the values of $f$ on the two levels, 
we interpolate $f$ from the coarser level onto the finer level using fourth 
order interpolation. If we compute the truncation error of multiple grid 
variables, we combine the error estimates for each variable at each point:
\begin{equation}
    \tau_{l}(\mathbf{i})
    =
    \sqrt{\sum_f \frac{\left(\tau_{l,f}(\mathbf{i})\right)^2}{L_f}}
    ,
    \label{eq:weighted-truncation-error}
\end{equation}
where $L_f$ is a normalizing factor for each variable $f$. We then set this as 
our tagging criterion in \eqref{eq:tagging-switch}: 
$C(\mathbf{i})=\tau_{l}(\mathbf{i})$. The free parameters in this scheme of 
tagging are the choice of grid variables that one computes truncation error 
estimates for and the normalization factors for each variable.

The main advantages of truncation error tagging are that it allows for a 
conceptually straightforward way to implement convergence tests in AMR codes: 
as one increases the base grid resolution, one should scale the truncation 
error tagging threshold for grid generation with the expected convergence of 
the code. Additionally, truncation error tagging is a ``natural'' tagging 
criterion as it refines regions that are most likely to be under resolved.

\subsection{Tagging criteria based on grid variables and derived quantities}
\label{sec:physics-tagging}

Some physical problems lend themselves to other tagging criteria, and \grchombo 
permits the user to easily specify refinement criteria based on any properties 
of the local grid variables or derived expressions of them, for example, 
derivatives or curvature scalars. We caution though that the tagging criteria 
we discuss below are not functions of geometric scalars, so the performance of 
a given criterion will depend on the formulation and gauge conditions used. 
Nevertheless, for the Bona-Masso-type slicing 
(\eqref{eq:Bona-Masso-slicing}) and gamma-driver 
(\eqref{eq:Gamma-driver-1}-\eqref{eq:Gamma-driver-2}) conditions we have 
the most experience with, these gauge conditions have proven to be reliable and 
robust.

First we discuss tagging criteria based on the conformal factor of the spatial 
metric $\chi$. Contours of $\chi$ can provide a good choice in dynamical BH 
cases as detailed in \sref{sec:black-hole-tagging} and  
\ref{sec:chihorizon}, to ensure that horizons are covered. Taking differences 
of $\chi$ across a cell using locally evaluated derivatives, i.e. using $C = 
\sqrt{\delta^{ij}\partial_i \chi \partial_j \chi} \Delta x$, also provides an 
efficient measure to refine key areas\footnote{Imposing simply that $\partial_i 
\chi$ (without the factor of $\Delta x$) is higher that some threshold results 
in unlimited regridding, since one does not reduce the local gradient in a 
variable by refinement, only the difference across the cell.}. In particular, 
using the second derivative of $\chi$, i.e. $C = 
\sqrt{\delta^{ij}\partial_i\partial_j \chi} \Delta x$, is efficient because 
usually it is the regions in which gradients are changing most rapidly that 
require greater resolution, rather than steep linear gradients. However, in 
practice any derivative can be used provided the thresholds are tuned 
appropriately for the problem at hand.

Alternatively, we find empirically that the sum of the absolute value of the 
different components of the Hamiltonian constraint proves to be an efficient 
tagging criteria in dynamical matter spacetimes. The condition is
\begin{equation}\label{eq:abs-Hamiltoniaon-constraint}
    C = \mathcal{H}_{\text{abs}} = |R| 
    + |\tilde{A}^{kl}\tilde{A}_{kl}|+\frac{2}{3}K^2+16\pi|\rho| +2|\Lambda|,
\end{equation}
where $R=\gamma^{ij}R_{ij}$ is the Ricci scalar. As we will see in 
\sref{sec:scalarstars}, this quantity generally remains constant in regions 
of spacetime where 
the individual 
metric and matter components oscillate in a stable, time-invariant manner 
(as in the case of the stable axion star we present later). Thus using 
this measure reduces the 
amount of spurious regridding that occurs, which in turn reduces errors 
introduced by that process. Where it starts to grow in some region, this 
generally reflects a decrease in the local dynamical timescales and thus 
physically motivated regridding.

A disadvantage of using these more arbitrary criteria over error tagging is 
that convergence testing is more challenging - one must ensure that similar 
regions are refined at the appropriate resolutions in the convergence runs, 
which necessitates tuning of the threshold at each different base resolution. 
Depending on the regridding condition, halving the threshold $\tau_R$, for 
example, may not result in double the resolution being applied. Nevertheless, 
if one ensures that the regions of most physical significance have an 
appropriate increase in refinement, convergence can usually be demonstrated, as 
we show below.

In order to obtain convergent results and use resolution most efficiently in a 
physical problem, it is often helpful to implement rules to enforce that given 
regions are refined for a given amount of time at least to a given 
level. Whilst this may seem to go against the spirit of AMR, it is easy to 
implement within that formalism as a secondary condition, and is often required 
to avoid excessive or insufficient tagging in very dynamical cases. For example, 
when one is not interested in resolving outgoing scalar radiation, one may 
choose to suppress regridding above a particular level outside of a particular 
radius. In the opposite sense, we often enforce extra regridding over the 
extraction surfaces for the Weyl scalars, to ensure that they have sufficient 
resolution and that noise is not introduced from grid boundaries crossing the 
spheres.

Several examples of the application of these criteria to black-hole binary 
inspirals and matter field evolutions are presented in the following, 
sections \ref{sec:bh} and \ref{sec:osc}. These two examples cover the main 
considerations when using AMR in NR codes. Spacetimes with singularities have 
particular requirements related to the resolution of the horizons. Furthermore, 
in dynamical matter spacetimes achieving an optimum frequency of regridding can 
be crucial in obtaining convergence. In the matter case we focus on an isolated 
real scalar (axion) star, which provides a very good test of AMR capabilities. 
In particular, it tests the ability to resolve stably oscillating matter 
configurations without excessive gridding and ungridding, and to follow the 
dynamical timescales of gravitational collapse, which are the key challenges in 
many simulations of fundamental fields in NR, including the modeling of cosmic 
strings, inflationary spacetimes and exotic compact objects. Fully AMR 
techniques are also likely to create significant challenges for high-resolution 
shock capturing, but we leave this topic for future investigations.


\section{Binary black-hole simulations with adaptive mesh refinement}
\label{sec:bh}

In this section, we demonstrate the efficacy of some of the techniques
discussed in \sref{sec:amr-techniques} in the context of BH binaries. 
To do this, we select
a representative sample of BH binary configurations, analyze the
accuracy of the resulting gravitational waveforms and compare the results 
obtained
with \grchombo to that obtained with a more conventional moving boxes style mesh
refinement code, \lean \cite{Sperhake:2006cy}. 

Before we present our results, we first provide details of the explicit tagging
criteria used in our \grchombo simulations and the methods we use to analyze
and compare our results.

\subsection{Methods}

\begin{figure}[t]
    \centering
    \includegraphics[width=0.5\columnwidth]{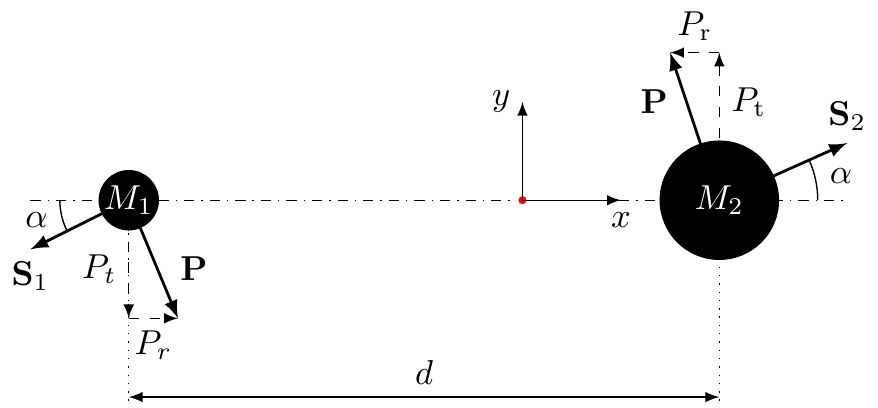}
    \caption{
    Schematic illustration of the parameters characterizing the
    BH binary configurations under consideration:
    the mass ratio $q=M_2/M_1>1$, 
    the initial separation $d$, the
    initial tangential linear momentum $P_{\mathrm{t}}$, the
    initial inward radial linear momentum $P_{\mathrm{r}}$,
    the dimensionless spin of each black hole $\chi_i=|\mathbf{S}_i|/M_i^2$
    and the angle of the spin in the orbital plane $\alpha$ relative to 
    the outward radial direction of the initial BH positions.
    }
    \label{fig:bh-schematic}
\end{figure}
\begin{table}[t]
    \lineup
    \centering
    \caption{\label{tab:BBHconfigs}
    A list of the parameter values (cf.~figure \ref{fig:bh-schematic}) 
    for the BH binary configurations simulated in this work.
    $M=M_1+M_2$ denotes the total black hole mass of the spacetime.
    }
    \begin{indented}
    \item[]
    \begin{tabular}{@{}rlllllll}
    \br
    Label & $q$ & $d/M$ & $P_{\mathrm{t}}/M$ & $P_{\mathrm{r}}/M$ & $\chi_i$ 
    & $\alpha$ & Reference
    \\
    \mr
    \texttt{q1-d12} & $1$   &12.21358 & 0.08417 & $5.10846\times 10^{-4}$ 
    & 0 & - & \cite{Khan:2019}
    \\
    \texttt{q2-d10} & $2$ & 10 & 0.08566 & 0 & 0
    & - & \cite{Radia:2021hjs}
    \\
    \texttt{q1-s09} & $1$ & 11.01768 & 0.075 & 0 & 0.9 & $30^{\circ}$ 
    & 
    \cite{Sperhake:2019wwo} 
    \\
    \br
    \end{tabular}
    \end{indented}
\end{table}
We consider three different BH binary configurations with the parameters 
provided in table \ref{tab:BBHconfigs} and illustrated schematically in 
figure \ref{fig:bh-schematic}. All simulations include an inspiral, merger and 
ringdown.

The first configuration consists of two equal-mass non-spinning BHs with a 
quasicircular inspiral lasting about 10 orbits; this configuration is labeled 
\texttt{q1-d12} (for mass ratio $q=1$ and distance $d\approx 12$). The 
parameters were computed in order to minimize the initial eccentricity of the 
simulation using standard techniques \cite{Khan:2019}.

The second configuration involves two BHs with mass ratio $2:1$. The inspiral 
is about 6 orbits and is approximately quasicircular. This is one of the 
configurations simulated in the \texttt{lq1:2} sequence of 
\cite{Radia:2021hjs}\footnote{Note that, in this paper, \grchombo was not 
used to evolve this configuration.}. Here we label this configuration 
\texttt{q2-d10} (for mass ratio $q=2$ and distance $d=10$).

The final configuration consists of a mildly eccentric inspiral of two 
equal-mass highly-spinning BHs. The spins lie in the plane as shown in 
figure \ref{fig:bh-schematic}, which is the ``superkick'' configuration 
\cite{Gonzalez:2007hi,Campanelli:2007cga,Campanelli:2007ew}. Here, the quantity 
we analyze is the gravitational recoil of the remnant BH. This configuration is 
taken from the sequence simulated in \cite{Sperhake:2019wwo} and we label 
it \texttt{q1-s09} (for mass ratio $q=1$ and spin $\chi=0.9$).

\subsubsection{GRChombo setup and tagging criteria}

For the \grchombo simulations of the BH binary configurations in 
table.~\ref{tab:BBHconfigs}, we use the CCZ4 equations 
(\ref{eq:CCZ4-chi}-\ref{eq:CCZ4-Gamma}) with the default damping parameters 
\eqref{eq:default-ccz4-damping} (note that in code units, $M=1$). We use the 
moving puncture gauge (\ref{eq:Bona-Masso-slicing}-\ref{eq:Gamma-driver-2}) 
with the default lapse parameters \eqref{eq:default-lapse-params} and the shift 
parameters $b_1=1,\,b_2=3/4$, $M\eta=1$ for \texttt{q1-d12} and \texttt{q1-s09} 
and $b_1=1,\,b_2=3/4$, $M\eta=3/4$ for \texttt{q2-d10}. For \texttt{q1-d12} and 
\texttt{q2-d10}, we use reflective BCs along one boundary to impose bitant 
symmetry (i.e.~symmetry across the equatorial plane) and Sommerfeld BCs for all 
other boundaries. Following sections \ref{sec:black-hole-tagging}, 
\ref{sec:asymmetric-grids-tagging} and \ref{sec:physics-tagging}, we use a 
tagging criterion of the following form
\begin{equation}
    C=\max\left(C_{\chi},C_{\text{punc}},C_{\text{ex}}\right),
    \label{eq:bh-tagging}
\end{equation}
where the quantities on the right-hand side are defined below. Note that we use 
the value $+\infty$ to denote a large value that always exceeds the threshold 
$\tau_R$.
\begin{enumerate}[label=(\roman*)]
    \item 
        $C_\chi$ tags regions in which the derivatives of the conformal
        factor $\chi$ become steep. It is the dominant criterion
        for the intermediate levels $l^{\max}_{\text{ex}}\leq l<\lmax-3$,
        where $l^{\max}_{\text{ex}}$ is the maximum extraction
        level (see item (iii) below).
        It is given by
        \begin{equation}
            C_{\chi} = \Delta x_l\sqrt{\sum_{i,j}\left(
            \partial_i\partial_j\chi\right)^2}\,,
        \end{equation}
        where $\Delta x_l$ is the grid spacing on refinement level $l$.
    \item
        $C_{\text{punc}}$ includes parts of the tagging criterion
        that use the location of the punctures. It is the
        dominant criterion on the finest three levels and is comprised of
        two parts, $C_{\text{insp}}$ and $C_{\text{merg}}$
        that are used depending on the coordinate distance
        between the punctures 
        $s_p=|\mathbf{x}_{\text{p},1}-\mathbf{x}_{\text{p},2}|$ 
        as follows:
        \begin{equation}
            C_{\text{punc}} =
            \begin{cases}
                C_{\text{insp}}, & s_p \geq M+b \\
                \max(C_{\text{insp}}, C_{\text{merg}}), 
                & 10^{-3} \leq s_p < M+b \\
                C_{\text{merg}}, & s_p < 10^{-3},
            \end{cases}
        \end{equation}
        where $M=M_1+M_2$ is the sum of the individual BH masses, 
        $b$ is a buffer parameter (cf. \sref{sec:black-hole-tagging}),
        and $10^{-3}$ is a choice in the cutoff for the distance between the 
        punctures $s_p$ which determines when the merger has completed.
        The inspiral criterion is given by
        \begin{equation}
            C_{\text{insp}}=
            \begin{cases}
            +\infty, & 
            \begin{array}{l}
                \text{if } \varrho_1 < (M_1+b) 2^{\min(\lmax-l-1, 2)},\\
                \text{or } \varrho_2 < (M_2+b) 2^{\min(\lmax-l-1, 2)},
            \end{array}\\
            0, & \text{otherwise},
            \end{cases}
        \end{equation}
        where $\varrho_i$ is the ``max'' or ``infinity'' norm
        \eqref{eq:box-tagging}
        of the coordinate position vector relative to
        puncture $i$. Similarly, the merger criterion is given by
        \begin{equation}
            C_{\text{merg}} =
            \begin{cases}
                +\infty, & \text{if }
                \varrho_{\centerofmass} < (M+b)2^{\min(\lmax-l-1,2)},\\
                0, & \text{otherwise},
            \end{cases}
        \end{equation}
        where $\varrho_{\centerofmass}$ is the max-norm \eqref{eq:box-tagging} 
        of the coordinate position vector relative to the center of mass
        $\mathbf{x}_{\centerofmass} 
        = (M_1\mathbf{x}_{p,1}+M_2\mathbf{x}_{p,2})/M$.
    \item
        $C_{\text{ex}}$ ensures the $\Psi_4$ extraction spheres are suitably
        well resolved. It is the dominant tagging criterion for 
        $0\leq l < l_{\text{ex}}^{\max}$
        and is given by $C_{\text{ex}}=\max_i \{C_{\text{ex},i}\}$,
        where $i$ labels the extraction spheres and
        \begin{equation}
            C_{\mathrm{ex},i} = 
            \begin{cases}
                +\infty, &\text{if } r < 1.2r_{\text{ex},i} \text{ and } 
                l < l_{\text{ex},i}, \\
                0, & \text{otherwise},
            \end{cases}
        \end{equation}
        where $r_{\text{ex},i}$ and $l_{\text{ex},i}$ are the radius and level 
        of the $i$th extraction sphere and $l^{\max}_{\text{ex}} = 
        \max_il_{\text{ex},i}$. The factor of $1.2$ is present to add a 
        $20\,\%$ buffer radius around the extraction spheres in order to 
        reduce the effect of spurious reflections off the refinement level 
        boundaries.
\end{enumerate}
A summary of the grid configuration parameters is given in 
table \ref{tab:grchombo-grids}.
\begin{table}[t]
    \lineup
    \centering
    \caption{\grchombo grid parameters for the configurations in 
    table \ref{tab:BBHconfigs} There are $(\lmax+1)$ refinement levels and the 
    coarsest level has length (without symmetries applied) $L$. The grid spacing 
    on the finest level is $\Delta x_{\lmax}$ and the minimum number of cells in 
    the buffer regions between consecutive refinement level boundaries is $n_B$.
    The regrid threshold for the tagging criterion \eqref{eq:bh-tagging} is 
    $\tau_R$ and the the BH buffer parameter is $b$.
    }
    \begin{indented}
    \item[]
    \begin{tabular}{@{}rllllll}
        \br
        Configuration & $\lmax$ & $L/M$ & $\Delta x_{\lmax}/M$ & $n_B$ 
        & $\tau_R$ & $b/M$ \\
        \mr
        \texttt{q1-d12} low & 9 & 1024 & $1/80$ & 20 & $0.016$ & $0.7$ 
        \\
        \texttt{q1-d12} medium & 9 & 1024 & $1/96$ & 24 & $0.0133$ & $0.7$ 
        \\
        \texttt{q1-d12} high & 9 & 1024 & $1/128$ & 32 & $0.01$ & $0.7$ 
        \\
        \texttt{q2-d10} low & 7 & \0512 & $1/88$ & 48 & $0.01$ & $0.467$ 
        \\
        \texttt{q2-d10} medium & 7 & \0512 & $1/104$ & 52 & $0.00923$ & $0.467$ 
        \\
        \texttt{q2-d10} high & 7 & \0512 & $1/112$ & 56 & $0.00857$ & $0.467$ 
        \\
        \texttt{q1-s09} low & 7 & \0512 & $1/64$ & 16 & $0.02$ & $0.7$ 
        \\
        \texttt{q1-s09} medium & 7 & \0512 & $1/96$ & 24 & $0.0133$ & $0.7$ 
        \\
        \texttt{q1-s09} high & 7 & \0512 & $1/112$ & 28 & $0.0114$ & $0.7$ 
        \\
        \br
    \end{tabular}
    \end{indented}
    \label{tab:grchombo-grids}
\end{table}

\subsubsection{Comparison code: Lean}
\label{sec:lean}

The \lean code~\cite{Sperhake:2006cy} is based on the \textsc{Cactus} 
computational toolkit~\cite{Goodale2002a} and uses the method of lines with 
fourth-order Runge-Kutta time stepping and sixth-order spatial stencils. The 
Einstein equations are implemented in the form of the 
Baumgarte-Shapiro-Shibata-Nakamura-Oohara-Kojima (BSSNOK) 
formulation~\cite{Nakamura:1987zz,Shibata:1995we,Baumgarte:1998te} with the 
moving-puncture gauge~\cite{Campanelli:2005dd,Baker:2005vv} (cf.~equations 
(\ref{eq:Bona-Masso-slicing}-\ref{eq:Gamma-driver-2})). The 
\textsc{Carpet} driver~\cite{Schnetter:2003rb} provides mesh refinement using 
the method of ``moving boxes.'' For the non-spinning binary configurations {\tt 
q1-d12} and {\tt q2-d10}, we use bitant symmetry to reduce computational 
expense, whereas configuration {\tt q1-s09} is evolved without symmetries. The 
computational domains used for these simulations are characterized by the 
parameters listed in table \ref{tab:lean-grids}. The domain comprises a 
hierarchy of $\lmax+1$ refinement levels labeled from $l=0,\ldots 
l_F,\ldots,\lmax$, with grid spacing given by \eqref{eq:grid-spacing}. 
Before applying the symmetry, for $l\leq l_F$ each level consists of a single 
fixed cubic grid of half-length $R_l=R_0/2^l$, and for $l_F<l\leq \lmax$, each 
level consists of two cubic components of half-length 
$R_l=2^{\lmax-l}R_{\lmax}$ centered around each BH puncture. We adopt this 
notation for consistency with that used to describe \grchombo. This translates 
into the more conventional \lean grid setup notation
(cf.~\cite{Sperhake:2006cy}) as
\begin{equation}
    \left\{(R_0,\ldots,2^{-l_F}R_0)
    \times  
    (2^{\lmax-l_F-1}R_{\lmax},\ldots,R_{\lmax}),\Delta x_{\lmax}\right\}.
\end{equation}
\begin{table}[t]
    \lineup 
    \centering
    \caption{\lean grid parameters for the configurations in 
    table \ref{tab:BBHconfigs}. There are $(\lmax+1)$ levels of which the 
    first $(l_F+1)$ comprises a single box that covers both BHs with the 
    remaining levels consisting of two separate box components that cover each 
    BH separately. The half-width of the coarsest level is $R_0$ and the 
    half-width of a single component on the finest level is $R_{\lmax}$. The 
    grid spacing on the finest level for the three resolutions used in the 
    convergence analysis is $\Delta x_{\lmax}$.
    }
    \begin{indented}
    \item[]
    \begin{tabular}{@{}rlllll}
        \br
        Configuration & $\lmax$ & $l_F$ & $R_0/M$ & $R_{\lmax}/M$ 
        & $\Delta x_{\lmax}/M$  
        \\
        \mr
        \texttt{q1-d12} & 9 & 5 & 512 & $1/2$ & $1/64,\, 1/96,\, 1/128$
        \\
        \texttt{q2-d10} & 8 & 4 & 256 & $1/3$ & $1/84,\, 1/96,\, 1/108$
        \\
        \texttt{q1-s09} & 8 & 3 & 256 & $1$ & $1/80,\, 1/88,\, 1/96$
        \\
        \br
    \end{tabular}
    \end{indented}
    \label{tab:lean-grids}
\end{table}
A CFL factor of $1/2$ is used in all simulations, and apparent horizons 
are computed with \textsc{AHFinderDirect}
\cite{Thornburg:1995cp,Thornburg:2003sf}.

For all our BH evolutions, with {\sc Lean} and {\sc GRChombo}, 
the initial data are constructed with the \textsc{TwoPunctures} spectral solver 
\cite{Ansorg:2004ds}.

\subsubsection{Gravitational wave analysis}
\label{sec:gw-analysis}

One of the most important diagnostics from our simulations is the GW signal 
which we compute from the Weyl scalar $\Psi_4$. For \grchombo, the calculation 
of $\Psi_4$ is explained in \ref{sec:psi4-appendix} and technical 
details of the extraction procedure can be found in 
\sref{sec:amrinterp-wave}. For \lean, details can be found in 
\cite{Sperhake:2006cy}. Below, we describe further analysis we have 
performed in order to compare the gravitational wave output from each code.

We start with the multipolar decomposition of the Weyl scalar,
\begin{equation}
    \Psi_4 
    = 
    \sum_{l=2}^\infty\sum_{m=-\ell}^\ell {}_{-2}Y^{\ell m}\psi_{\ell m}.
\end{equation}
Next, we translate to the gravitational-wave strain $h$ according to
\begin{equation}
  \Psi_4 = \ddot{h} = -\ddot{h}^{+} + \iu \ddot{h}^{\times} 
\end{equation}
which gives us the strain multipoles as $\ddot{h}^+_{\ell m}=-{\textbf{\sf 
Re}(\psi_{\ell m})}$ and $\ddot{h}^{\times}_{\ell m}={\textbf{\sf 
Im}(\psi_{\ell m})}$. To avoid spurious drift resulting from numerical 
inaccuracies, we perform the necessary integrations in time in the Fourier 
domain \cite{Reisswig:2010di}. We then rewrite the strain modes in terms of 
their amplitude and phase
\begin{equation}\label{eq:strain-modes-amp-phase}
    -h^+_{\ell m}+\iu h^{\times}_{\ell m}=h^A_{\ell m}
    \exp\left(\iu h^{\phi}_{\ell m}\right),
\end{equation}
where multiples of $2\pi$ are added to $h^{\phi}_{\ell m}$ appropriately in 
order to minimize the difference between consecutive data points.

The \emph{radiated} quantities derived from $\Psi_4$ are affected by two main 
error sources; the discretization error due to finite resolution and an 
uncertainty arising from the extraction at finite radii instead of null 
infinity. We determine the former by conducting a convergence analysis of the 
quantities extracted at finite radius. In order to determine the second error 
contribution, we compute a given radiated quantity $f$ at several finite 
extraction radii and extrapolate to infinity by fitting a polynomial in $1/r$ 
of the form
\begin{equation}
    f_N(u,r)=\sum_{n=0}^{N}\frac{f_{n,N}(u)}{r^n}.
\end{equation}
Here, $r$ is the coordinate radius and $u=t-r^{\ast}$ denotes the retarded time 
evaluated with the tortoise coordinate
\begin{equation}
    r^\ast = r + 2M \ln \left|\frac{r}{2M}-1\right|.
\end{equation}
We uniformly observe that time shifts in terms of $r^{\ast}$ result in slightly 
better alignment of wave signals extracted at different coordinate radii $r$. 
If we take $f_{0,N}(u)$ as our estimate of the extrapolated quantity, we then 
estimate the error $\epsilon$ in our result from $r=r_{\text{ex}}$ by computing
\begin{equation}
    \epsilon_{f,r_{\text{ex}},N}=\left|f(u,r_{\text{ex}})-f_{0,N}(u)\right|.
\end{equation}
Typically, and unless stated otherwise, we set $N=1$ and drop the $N$ 
subscripts. Our total error budget is then given by the sum of the 
discretization and extraction uncertainties.

We quantify the agreement between the two codes' results in the context of GW 
analysis by computing the overlap following the procedure of 
\cite{Damour:1997ub,Miller:2005qu}. In the following, we restrict our 
analysis to the dominant (2,2) quadrupole part of the signal and drop the 
subscript ``$\ell=2, m=2$''. Before computing the overlap, we extrapolate the 
strain to infinity using the procedure explained above.

Given the power spectral density $S_n(f)$ of a detector's strain noise as a 
function of frequency $f$, the inner product of two signals $g$, $h$ on the 
space of waveforms is given by\footnote{We use in our calculation one-sided, as 
opposed to two-sided, spectral power densities, i.e.~we only consider 
non-negative frequencies, hence the factor $4$ in (\ref{eq:innerproduct}).}
\begin{equation}
  \langle g | h\rangle := 4\textbf{\sf Re}\left\{
  \int_0^{\infty} \frac{\tilde{g}^*(f)\tilde{h}(f)}{S_n(f)}\,\rmd f
  \right\},
  \label{eq:innerproduct}
\end{equation}
where the Fourier transform is defined by
\begin{equation}
    \tilde{g}(f):=\int^\infty_{-\infty}g(t)\mathrm{e}^{-2\pi\iu ft}\,\rmd t.
\end{equation}
We next define the overlap of the two signals as the normalized inner product 
maximized over shifts $\Delta t$, $\Delta h^\phi$ in time and phase,
\begin{equation}
  \rho(g,h) := \max_{\Delta h^\phi, \Delta t}\frac{\langle g|h\rangle}
  {\sqrt{\langle g|g\rangle\langle h|h\rangle}}\,.
  \label{eq:overlap}
\end{equation}
The quantity $1-\rho(g,h)$ then provides a measure for the \emph{discrepancy} 
between the two waveforms, analogous to the {\it mismatch} introduced as a 
measure for signal-to-noise reduction due to model imperfections in GW data 
analysis \cite{Owen:1995tm,Lindblom:2008cm}.

For \texttt{q1-s09}, we instead analyze the convergence of the linear momentum 
radiated in GWs in the form of the BH recoil velocity or kick. First we compute 
the radiated momentum $\mathbf{P}^{\text{rad}}$ using equation (7) of 
\cite{Radia:2021hjs} and then compute the recoil velocity--which must lie 
in the $z$-direction by symmetry--using $v=-P^{\text{rad}}_z/M_{\text{fin}}$. 
Since the radiated momentum can be written in terms of a sum, with each term 
involving several multipolar amplitudes $\psi_{\ell m}$ (equation (40) in 
\cite{Ruiz:2007yx}), analyzing this quantity has the benefit of 
additionally indirectly comparing the agreement of higher order multipoles 
(i.e. $\ell>2$) between the codes.

\subsection{Results}

\begin{figure}[t]
    \centering
    \noindent\makebox[\textwidth]{%
    \subfloat[]{\label{fig:q1-convergence-grchombo}
    \includegraphics[width=0.6\columnwidth]{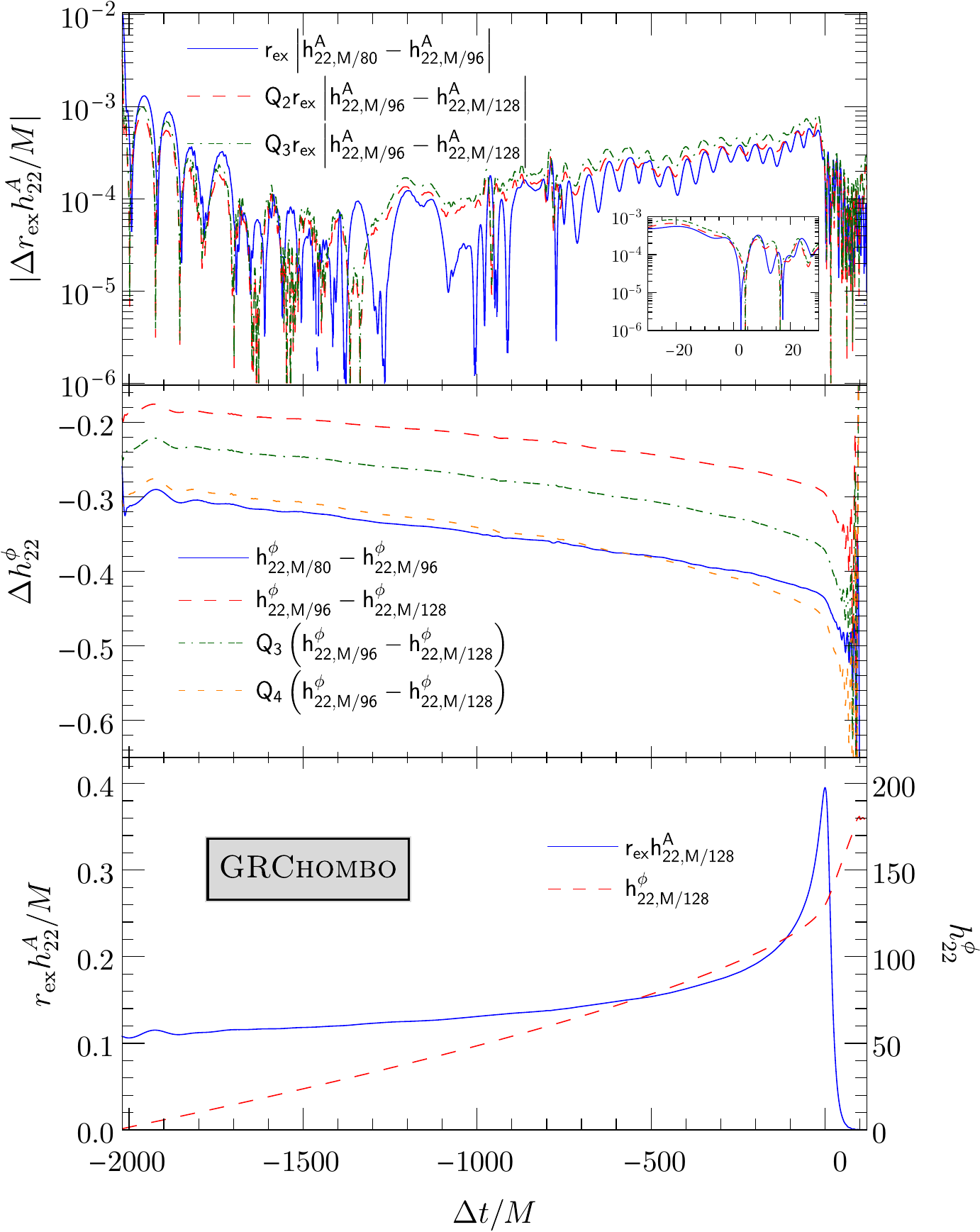}
    }
    \hfill
    \subfloat[]{\label{fig:q1-convergence-lean}
    \includegraphics[width=0.6\columnwidth]{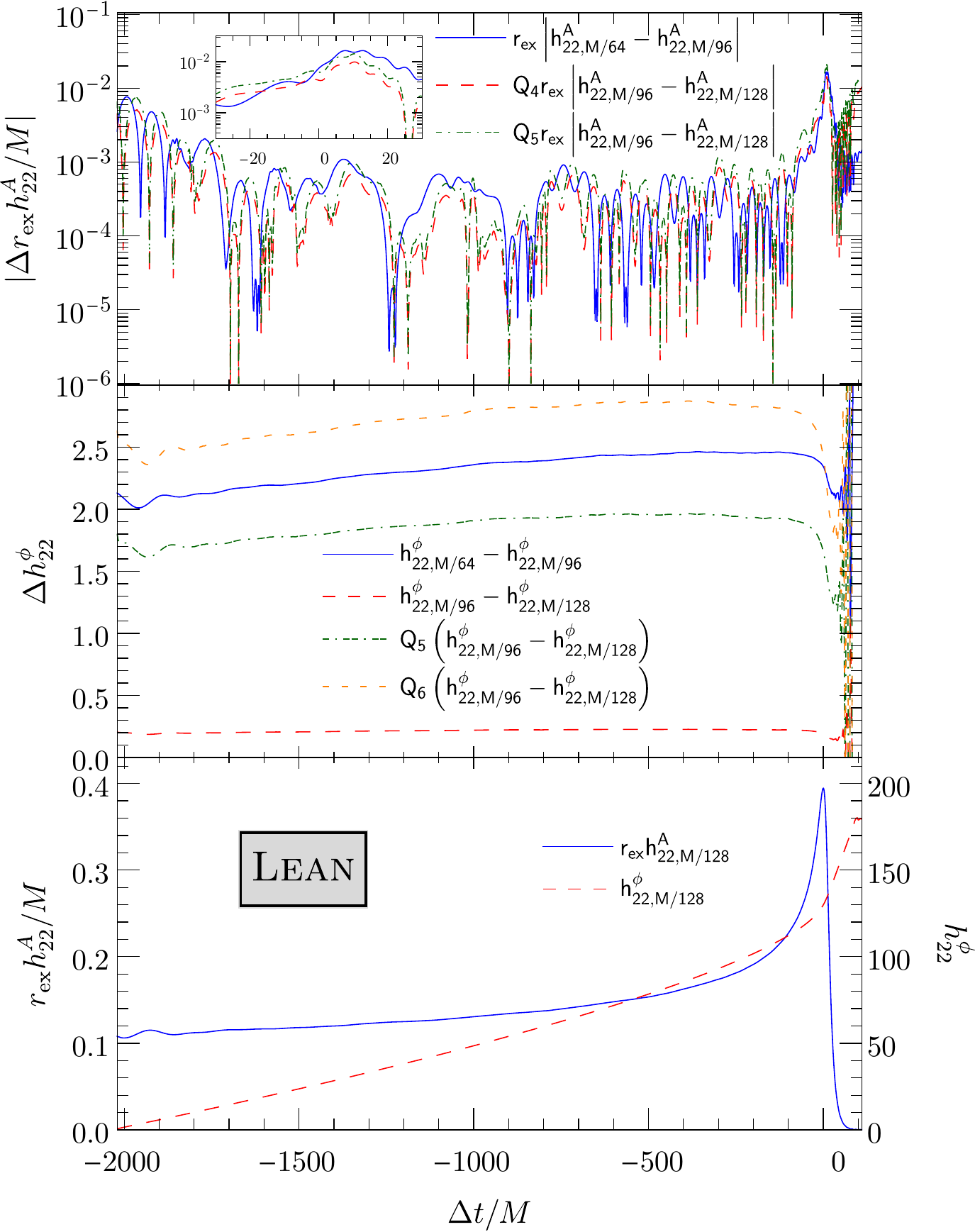}
    }
    }
    \caption{Convergence
    of the quadrupole mode of the strain $h_{22}=-h^+_{22}+\iu h^{\times}_{22}$ 
    calculated from the values of $\Psi_4$ extracted for configuration
    \texttt{q1-d12} at $r_{\mathrm{ex}}=120\,M$ for both \grchombo with finest 
    grid resolutions $\Delta x_{\lmax}=M/80$, $M/96$ and  $M/128$ 
    \protect\subref{fig:q1-convergence-grchombo} and \lean with finest grid 
    resolutions $\Delta x_{\lmax}=M/64$, $M/96$ and $M/128$ 
    \protect\subref{fig:q1-convergence-lean}.
    \textit{Top panels}:
    Convergence of the amplitude $h^A_{22}=|h_{22}|$. 
    The difference between the higher resolution results is rescaled according 
    to fourth and fifth order convergence for \lean
    and according to second and third-order convergence for \grchombo.
    In each case, the inset shows an interval around the peak amplitude.
    \textit{Middle panels}: 
    Convergence of the phase $h^{\phi}_{22}=\mathrm{Arg}(h_{22})$. The 
    difference between the higher resolution results is rescaled according to
    fifth and sixth order convergence for \lean and third and fourth order 
    convergence for \grchombo.
    \textit{Bottom panels}: 
    For reference we plot the amplitude $h^A_{22}$ and the phase 
    $h^{\phi}_{22}$ of the highest resolution waveform on the same time axis.
    For the two lower resolution waveforms from each code, we have time-shifted 
    each of them in order to maximize the overlap (cf.~\eqref{eq:overlap}) 
    with the highest resolution waveform. $\Delta t=0$ corresponds to the 
    maximum in $h^A_{22}$ for the highest resolution waveform.
    }
    \label{fig:q1-convergence}
\end{figure}

For each configuration in table~\ref{tab:BBHconfigs}, we have performed three 
simulations at different resolutions with both \grchombo and \lean in order to 
calibrate their accuracy which we discuss below. The respective grid 
configurations are given in tables \ref{tab:grchombo-grids} and 
\ref{tab:lean-grids}.

For the first configuration \texttt{q1-d12} of an equal-mass binary, we show 
the convergence analysis in figure \ref{fig:q1-convergence} with the analysis 
for \grchombo on the left and for \lean on the right. For \lean, we observe 
convergence of about fourth order in the amplitude and between fifth and sixth 
order in the phase of the quadrupole mode $h_{22}$ of the strain 
\eqref{eq:strain-modes-amp-phase}. For \grchombo we observe convergence of 
about second order in the amplitude and about fourth order in the phase of the 
same mode. We note that, as mentioned in \cite{Radia:2021hjs}, higher 
resolutions were required with \grchombo in order to enter the convergent 
regime.

By comparison with a Richardson extrapolation, we estimate the discretization 
errors in the amplitude and phase of the finest resolution simulations from 
both codes as follows. Excluding the early parts of the signal dominated by 
``junk'' radiation and the late part of the ringdown which is dominated by 
noise, we obtain a discretization error of $\Delta h^A_{22}/h^A_{22} \lesssim 
1\,\%$ in the amplitude assuming fourth order convergence for \lean and second 
order convergence for \grchombo. Up to the late ringdown where the phase 
becomes inaccurate, we estimate the phase error is $\Delta h^{\phi}_{22} \le 
0.15$ assuming sixth order convergence for \lean and fourth order convergence 
for \grchombo.

Following the procedure in \sref{sec:gw-analysis}, we estimate the error in 
the phase, due to finite-radius extraction, is 
$\epsilon_{h^{\phi}_{22},120M}\lesssim 0.4$, and, in the amplitude is 
$\epsilon_{h^A_{22},120M}/h^A_{22}\lesssim 8\,\%$ (although this steadily 
decreases towards $\lesssim 2\%$ near merger) for both codes. Here we have 
ignored the early part of the signal where the amplitude is dominated by the 
``junk'' radiation up to $u=300M$.

\begin{figure}[t]
    \centering
    \includegraphics[width=0.7\columnwidth]{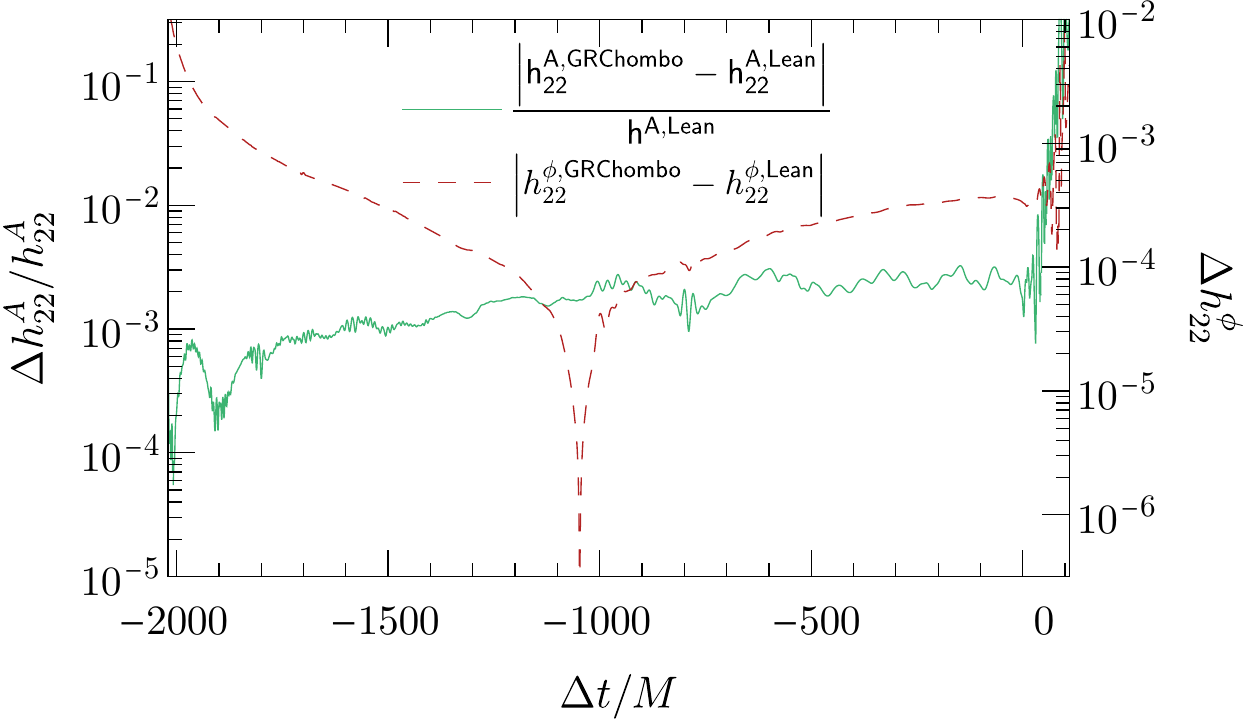}
    \caption{The relative and absolute difference between the \grchombo and 
    \lean outputs for the amplitude and phase of $h_{22}$ respectively from the 
    simulation of configuration \texttt{q1-d12}. In both cases, the data comes 
    from the simulations with finest grid spacing $\Delta x_{\lmax}=M/128$ with 
    $\Psi_4$ extracted at $r_{\text{ex}}=120M$. As for the convergence plots in 
    figure \ref{fig:q1-convergence}, the time has been shifted in order to 
    maximize the overlap (cf.~\eqref{eq:overlap}) between the two waveforms 
    and $\Delta t=0$ at the peak in $h^{A,\lean}_{22}$.
    }
    \label{fig:q1-comparison}
\end{figure}

We next directly compare the results of the two codes by computing the relative 
difference in the amplitude $h^A_{22}$ and the absolute difference in the phase 
$h^{\phi}_{22}$ which is shown in figure \ref{fig:q1-comparison}. Again, 
ignoring the early part of the signal and the late ringdown, the relative 
difference in the amplitudes is $\lesssim 1\,\%$, consistent with the individual 
error estimates from the two codes. The discrepancy in phase remains 
$\mathcal{O}(10^{-3})$ or smaller throughout the inspiral, merger and early 
ringdown--well within the error estimates of each code.

For the first asymmetric BH binary configuration, \texttt{q2-d10}, we proceed 
in the same way. We study the convergence in analogy to 
figure \ref{fig:q1-convergence}. Ignoring again the contamination at early 
times, we obtain third-order convergence in the amplitude and fifth-order 
convergence in the phase for \lean. For \grchombo, we obtain fourth order 
convergence in the amplitude and mild overconvergence of about eighth order in 
the phase\footnote{We assume fourth order convergence for our \grchombo phase 
error.}. This leads to uncertainty estimates of $\Delta 
h^A_{22}/h^A_{22}\lesssim 2.5\,\%$ in the amplitude and $\Delta h^{\phi}_{22} 
\lesssim 0.25$ in the phase for both codes.

The error due to finite-radius extraction in the amplitudes is 
$\epsilon_{h^A_{22},86.7M}/h^A_{22}\lesssim 10\,\%$ in the early inspiral 
decreasing down to $\lesssim 2\,\%$ in the late inspiral, and in the phase is 
$\epsilon_{h^\phi_{22},86.7M} \sim 0.5$ for both codes.

\begin{figure}[t]
    \centering
    \includegraphics[width=0.7\columnwidth]{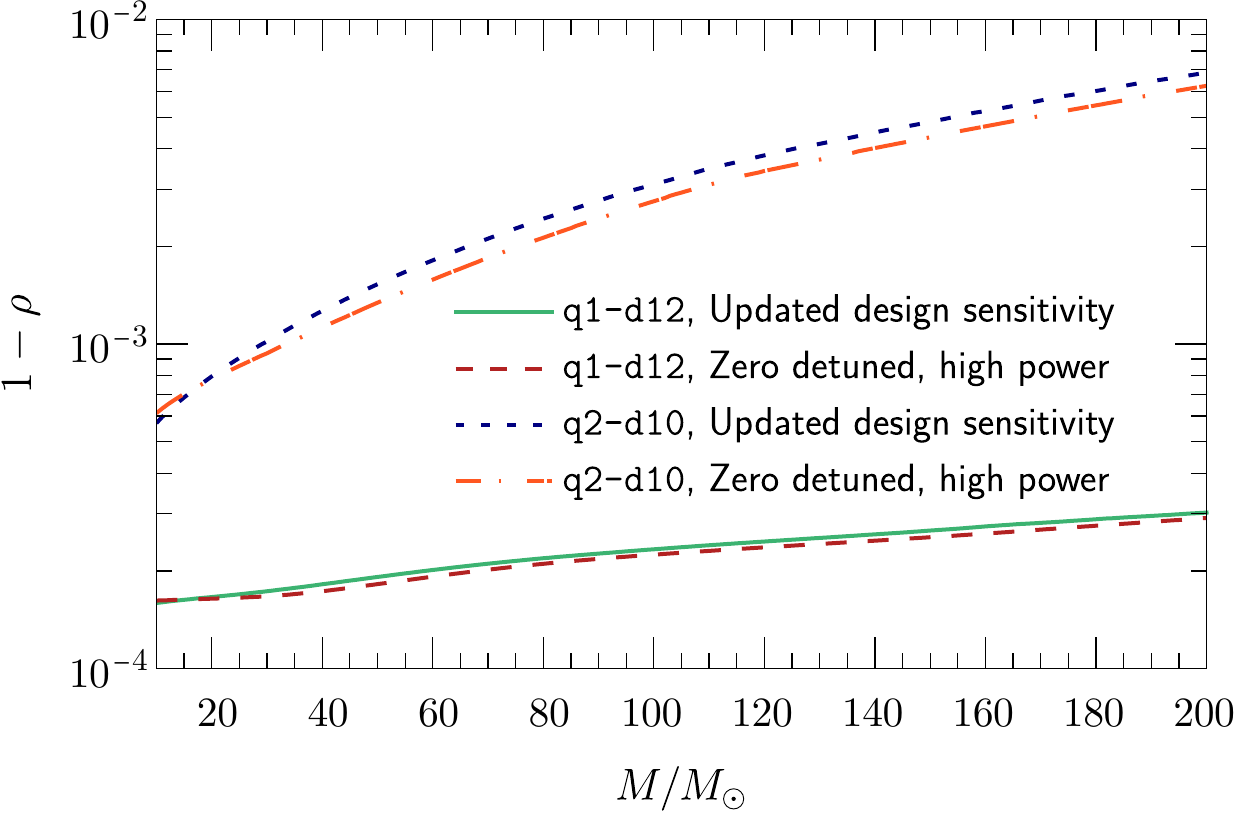}
    \caption{The discrepancy $1-\rho$ between the $(\ell,m)=(2,2)$
    mode of the gravitational wave signal from the
    \texttt{q1-d12} (10 orbits, non-spinning, equal mass)
    and \texttt{q2-d10} (6 orbits, non-spinning, $2:1$ mass ratio)
    BH binary configurations simulated with \lean and \grchombo.
    For \texttt{q1-d12}, we use the simulation with resolution
    $\Delta x_{\lmax}=M/128$ for both codes and for \texttt{q2-d10}, we use
    the simulation with resolution $\Delta x_{\lmax}=M/96$ for both codes.
    For each configuration, we show the difference computed
    with the updated Advanced LIGO sensitivity design curve 
    (\texttt{aLIGODesign.txt} in \cite{aLIGOupdatedantsens})
    and the zero detuned, high power noise curve from the 
    Advanced LIGO anticipated sensitivity curves 
    (\texttt{ZERO\_DET\_high\_P.txt} in \cite{aLIGOantsens}).
    }
    \label{fig:q1_mismatch}
\end{figure}

In figure \ref{fig:q1_mismatch}, we display as a function of the total mass $M$ 
the discrepancy $1-\rho$ (where $\rho$ is the overlap given by 
\eqref{eq:overlap}) between the \grchombo and \lean waveforms for both 
\texttt{q1-d12} and \texttt{q2-d10}, with the spectral noise density $S_n(f)$ 
given by (i) the updated Advanced LIGO sensitivity design curve 
(\texttt{aLIGODesign.txt} in \cite{aLIGOupdatedantsens}) and (ii) the zero 
detuned, high power noise curve from the Advanced LIGO anticipated sensitivity 
curves (\texttt{ZERO\_DET\_high\_P.txt} in \cite{aLIGOantsens}).
For \texttt{q1-d12}, the figure demonstrates excellent agreement of the two 
waveforms for the entire range $M = 10\ldots 200\,M_{\odot}$ with a discrepancy 
$1-\rho\ \approx 0.03\,\%$ or less, whereas for \texttt{q2-d10}, the agreement 
is not quite as strong but nevertheless demonstrates very good consistency with 
a discrepancy $1-\rho \approx 0.7\,\%$ or less. The larger difference for 
\texttt{q2-d10} compared to \texttt{q1-d12} may be attributed to the slightly 
lower resolutions employed for this configuration, especially near the smaller 
BH. To put these numbers into context, Lindblom\textit{et 
al.}~\cite{Lindblom:2008cm} estimate that a mismatch of $3.5\,\%$ would result 
in a reduction in the GW event detection rate by about 10\,\%.

\begin{figure}[t]
    \centering
    \noindent\makebox[\textwidth]{%
    \subfloat[]{\label{fig:superkick-convergence-grchombo}
    \includegraphics[width=0.55\columnwidth]{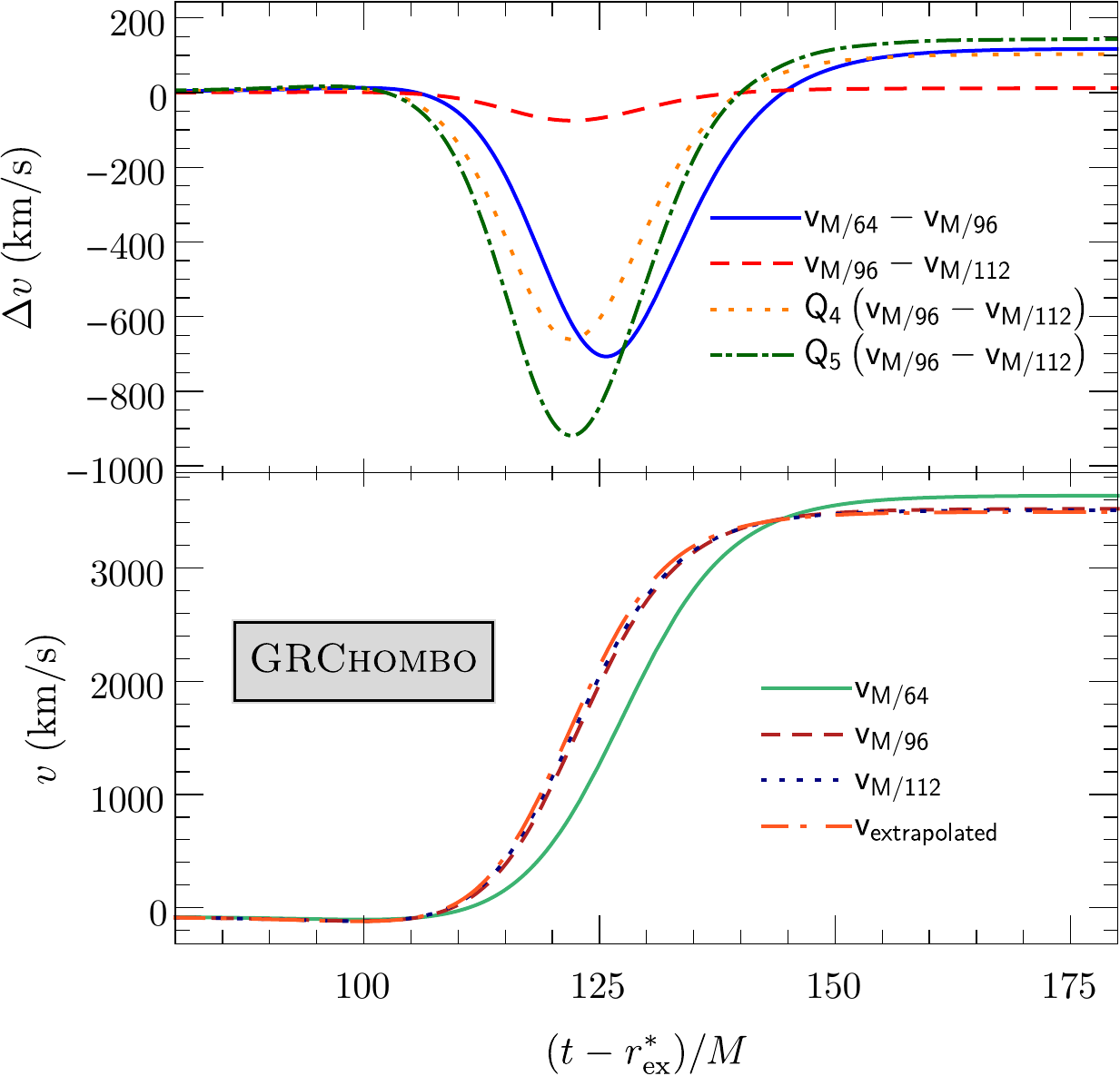}
    }
    \hfill
    \subfloat[]{\label{fig:superkick-convergence-lean}
    \includegraphics[width=0.55\columnwidth]{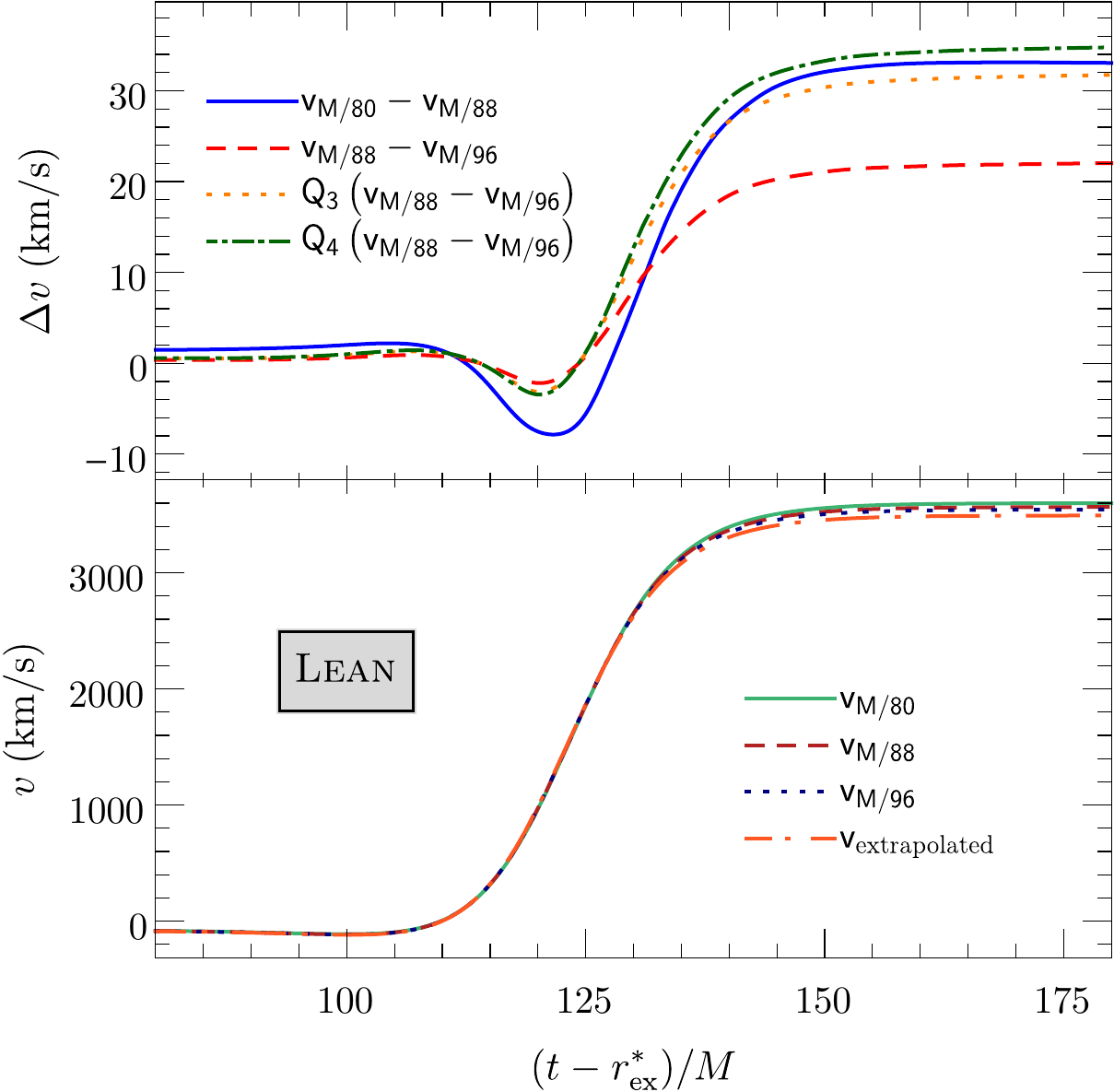}
    }
    }
    \caption{Convergence plots for the accumulated linear momentum radiated from 
    configuration \texttt{q1-s09} for \grchombo with finest grid 
    resolutions $\Delta x_{\lmax}=M/64$, $M/96$ and $M/112$ 
    \protect\subref{fig:superkick-convergence-grchombo} and for 
    \lean with finest grid resolutions $\Delta x_{\lmax}=M/80$, $M/88$ and 
    $M/96$ \protect\subref{fig:superkick-convergence-lean}. This is 
    shown in the form of the BH recoil velocity in the bottom panels.
    For both codes, the radiated linear momentum is calculated from the 
    extracted $\Psi_4$ values at $r_{\text{ex}}=90M$, and the extrapolated curve 
    corresponds to a Richardson extrapolation assuming fourth order convergence. 
    In the top panels, we show the difference between the results from different 
    resolutions with rescalings according to third and fourth order convergence 
    for \lean and according to fourth and fifth order convergence for \grchombo.
    }
    \label{fig:superkick-convergence}
\end{figure}

\begin{figure}[t]
    \centering
    \includegraphics[width=0.62\columnwidth]{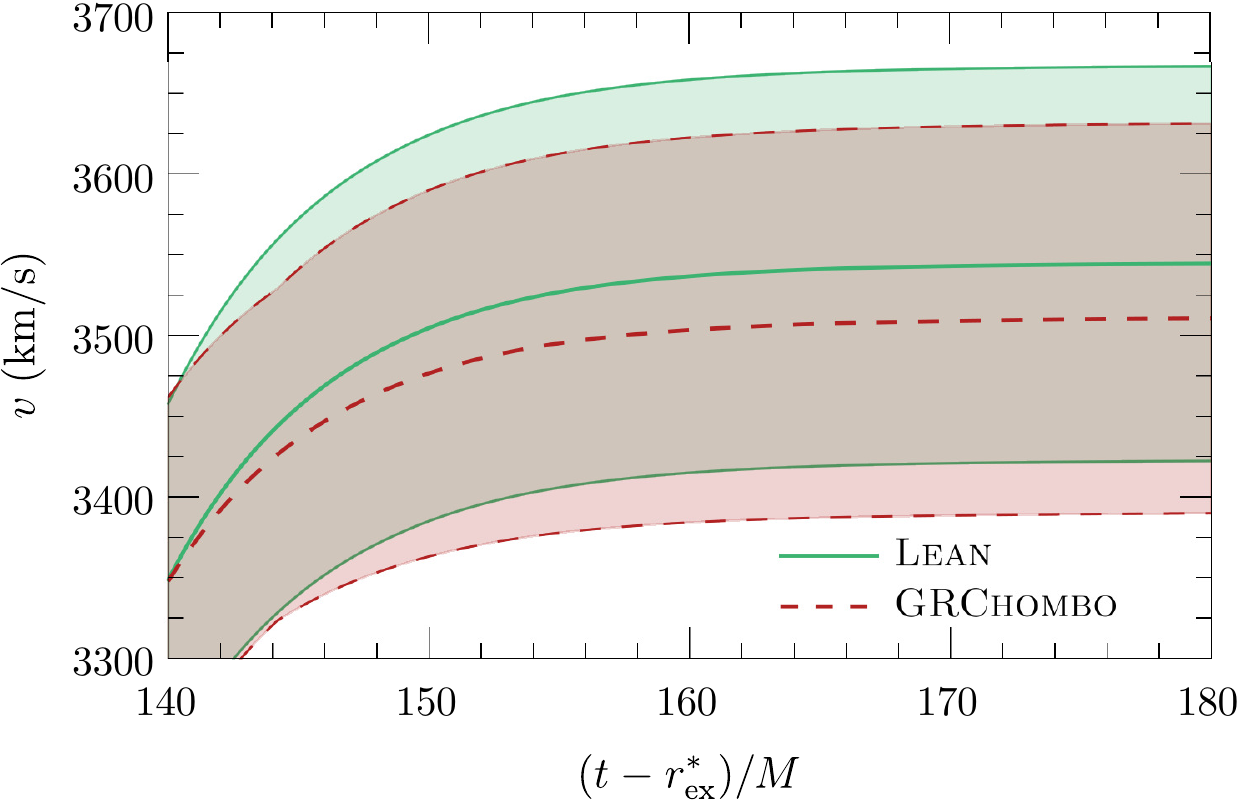}
    \caption{The accumulated radiated linear momentum at the end of the highest
    resolution simulations of configuration \texttt{q1-s09} from each code. 
    The linear momentum is shown in the form of the BH recoil velocity and the 
    error bands show our estimate of the total error coming from both 
    discretization and finite-radius effects.}
    \label{fig:superkick-comparison}
\end{figure}

Our final BH binary features asymmetry in the form of non-zero spins. This 
time, we focus on the BH recoil velocity $v$ calculated from the linear 
momentum radiated in GWs, and the analysis is shown in 
figure \ref{fig:superkick-convergence}.
From the plots, we can see that \lean exhibits convergence between third and 
fourth order, whilst \grchombo exhibits convergence between fourth and fifth 
order. We illustrate our estimate of the total error for each code in 
figure \ref{fig:superkick-comparison}. Here, the error bands---around the curve 
from the highest resolution simulation in each case---correspond to the 
difference with the Richardson extrapolated curve assuming fourth-order 
convergence plus the estimated error due to finite-radius extraction\change{}{
(about $1.5\,\%+2\,\%$ for \lean and $0.5\,\%+3\,\%$ for \grchombo)}. This 
total error is about $3.5\,\%$ for both codes.

\change{}{As is not uncommon, the convergence orders obtained from numerical
relativity simulations can be fickle due to the various ingredients in the codes
with differing orders of accuracy which can dominate in certain regimes.
This inherent complexity makes it difficult to attribute the difference in 
convergence orders we obtain between the two codes and we therefore do not 
attempt to do so.}


\section{Comparing tagging criteria using 
axion stars}
\label{sec:osc}

In order to demonstrate the application of our techniques to problems with 
matter fields and dynamically varying length scales, we consider the evolution 
of a single axion star---a compact object composed of a real scalar bosonic 
field. We analyze the evolution of a star that is stable on the timescale of the 
simulation, as well as one in which the self-interaction is increased in order 
to trigger gravitational collapse to a BH. As discussed previously, this simple 
example tests many of the key requirements in using AMR to evolve fundamental 
fields coupled to gravity, in particular, the ability to follow stable 
oscillations and to adapt to changing dynamical timescales. Similar 
considerations apply, for example, to cosmological spacetimes, cosmic strings 
and collisions of exotic compact objects.

We demonstrate the use of two effective tagging criteria; first, tagging by the 
magnitude of terms in the Hamiltonian constraint \eqref{eq:Habs}, and second, by 
the numerical truncation error between refinement levels 
\eqref{eq:weighted-truncation-error}.

\subsection{Methods}
\label{sec:osc_methods}

\subsubsection{Setup}
\label{sec:osc_setup}

\begin{table}[t]
    \lineup
    \centering
    \caption{\grchombo grid parameters for axion star configurations using 
    different tagging criteria. There are $(\lmax+1)$ refinement levels and the 
    coarsest level has length (without symmetries applied) $L$. The grid spacing 
    on the coarsest level is $\Delta x_0$ and the minimum number of cells in 
    the buffer regions between consecutive refinement level boundaries is $n_B$.
    The regrid thresholds for the different tagging criteria are given by 
    $\tau_R$. We consider two cases; a stable axion star ($f_a = 1$) and an 
    unstable collapse to a BH ($ f_a = 0.05$), with $\mu = m_a c / \hbar = 1$ 
    in code units.}
    \begin{indented}
    \item[]
    \begin{tabular}{@{}lllllll}
        \br
        $f_a$ & Tagging & $\lmax$ & $\mu L$ & $\mu\Delta x_0$ & $n_B$ & $\tau_R$ 
        \\
        \mr
        1.0 & Ham & 3 & \0512 & \0\04 & \08 & $0.1$ 
        \\
        1.0 & Ham & 3 & \0512 & \0\02 & 16 & $0.05$ 
        \\
        1.0 & Ham & 3 & \0512 & \0\01  & 32 & $0.025$ 
        \\
        1.0 & Trunc & 3 & \0512 & \0\04 & \08 & $0.0625$ 
        \\
        1.0 & Trunc & 3 & \0512 & \0\02 & 16 & $3.91\,\rm{x}\,10^{-3}$ 
        \\
        1.0 & Trunc & 3 & \0512 & \0\01 & 32 & $2.44\,\rm{x}\,10^{-4}$ 
        \\
        \mr
        0.05 & Ham & 8 & 1024 & \0\04 & \08 & $0.1$ 
        \\
        0.05 & Ham & 8 & 1024 & 2.67 & 12 & $0.067$ 
        \\
        0.05 & Trunc & 8 & 1024 & \0\04 & \08 & $0.001$ 
        \\
        0.05 & Trunc & 8 & 1024 & 2.67 & 12 & $2.96\,\rm{x}\,10^{-4}$ 
        \\
        \br
    \end{tabular}
    \end{indented}
    \label{tab:grchombo-grids-oscillaton}
\end{table}

We consider the evolution of two different axion star configurations. Axion 
stars are quasi-equilibrium configurations of a self-gravitating real scalar 
field $\phi$ \cite{Alcubierre:2003sx} that is subject to a periodic 
self-interaction potential $V(\phi)$. A canonical potential is
\begin{equation}
    V(\phi) = \mu^2f^2_a[1 - \cos{(\phi/f_a)}]\,,
    \label{eqn:axion_potential}
\end{equation} 
where this form arises as a result of the spontaneously broken $U(1)$ 
Peccei-Quinn symmetry and subsequent ``tilting'' of the potential due to 
instanton effects \cite{Peccei:1977hh,Weinberg:1977ma}. The decay constant 
$f_a$ quantifies the symmetry breaking scale and determines the strength of the 
scalar field self-interactions (their strength for a given central amplitude is 
inversely related to $f_a$) and $\mu = m_a c / \hbar$ is an inverse length 
scale related to the scalar mass\footnote{In Planck units one can write $\mu = 
m_a/M_{\text{Pl}}^2$, where $M_{\text{Pl}}$ is the Planck mass.} $m_a$. Axion 
stars on the main stability branch are characterized by their central amplitude 
$\phi_0$ or equivalently their ADM mass $M_{\rm ADM} \sim \mu^{-1}$. They have 
a physical size $R$ (defined as the radius containing $99\,\%$ of the total 
mass) that is approximately inversely related to their ADM mass, and thus a 
useful descriptor is their compactness $\mathcal{C} = M_{\rm ADM}/R$. Axion 
stars with $\mathcal{C}\sim 1/2$ are highly relativistic and may form BHs if 
they collapse or collide. For $m_a \sim 10^{-14}$ eV, they are of a mass and 
size comparable to solar mass BHs, and thus potentially of astrophysical 
interest. Further details related to the setup used here can be found in 
\cite{Clough:2018exo, Clough:2021qlv}, and a useful general review of 
axion physics is provided in \cite{Marsh:2015xka}.

The equation of motion for the scalar field $\phi$ is given by the Klein-Gordon 
equation for a real scalar field minimally coupled to gravity
\begin{equation}
    \nabla_\mu \nabla^\mu \phi - \frac{dV}{d\phi} = 0\, ,
    \label{eqn:EKG}
\end{equation}
and the system is completed with the Z4 equations \eqref{eq:Z4-covariant} for 
the metric components. To construct localised, quasi-equilibrium oscillatory 
(axion star) solutions, we solve the Einstein-Klein-Gordon (EKG) system of 
equations with a harmonic field ansatz and appropriate boundary conditions 
\cite{Urena-Lopez:2002ptf, Alcubierre:2003sx}. 

Unlike the case of complex scalar boson stars, for axion star solutions the 
metric components $g_{\mu\nu}$ also oscillate in time, with energy being 
transferred between the matter and curvature terms in the Hamiltonian 
constraint \cite{Clough:2021qlv}. This makes them challenging targets for 
dynamical refinement; simple criteria based solely on matter field gradients 
will fail to achieve a stable grid structure, as the gradients change over time 
even in the quasi-stable case. If the gradients are close in value to a tagging 
threshold, frequent regridding will occur, which introduces errors. 

The stability of the axion star solution comes from the balance of its tendency 
to disperse due to gradient pressure from spatial field derivatives, with the 
tendency to collapse due to its energy density. The relative strengths of 
these effects determines whether the 
axion star remains stable, disperses through scalar radiation or collapses to a 
black hole when perturbed. In particular, if the self-interaction scale $f_a$ is 
too low, this can cause the axion star to collapse to a BH 
\cite{Helfer:2016ljl}. In the last case, a key AMR challenge is determining 
tagging criteria that progressively track the axion star collapse without 
triggering too frequent regridding from the more rapid field oscillations. 

We consider two cases, both with central amplitude $\phi_0 = 0.020$ and $\mu 
M_{\rm ADM}=0.4131$:
\begin{enumerate}[label=(\roman*)]
    \item An axion star with weak self-interactions ($f_a=1$)\footnote{The 
    decay constant $f_a$ is dimensionless in geometric units as used here; to 
    obtain its value in Planck units one simply multiplies by the Planck mass 
    $M_{\text{Pl}}$.}, where the scalar field and metric oscillate over time in 
    a localized configuration that is stable over time periods much longer than 
    that of the simulation. We would ideally like the refinement to remain 
    constant, despite the oscillations of the fields.
    \item An unstable configuration where we increase the attractive self 
    interaction by reducing the self interaction scale to $f_a=0.05$, such that 
    the axion star is destabilized and undergoes collapse to a BH. We need the 
    mesh refinement to follow this process sufficiently rapidly, but without 
    excessive regridding.
\end{enumerate}

To evolve this system in \grchombo, the EKG equation \eqref{eqn:EKG} is 
decomposed into two first order equations in the 3+1 formulation
\begin{align}
    \partial_t\phi &= \beta^i\partial_i\phi + \alpha\Pi,\\
    \partial_t\Pi &= \beta^i\partial_i\Pi + \alpha \gamma^{ij} 
    (\partial_i\partial_j\phi + 
    \partial_i\phi\partial_j\alpha) 
    + \alpha\left(K\Pi - \gamma^{ij}\Gamma^k_{ij}\partial_k\phi
    - \frac{dV}{d\phi}\right),
\end{align}
and added to the CCZ4 evolution scheme (\ref{eq:CCZ4-chi}-\ref{eq:CCZ4-Gamma}). 
The initial data are set up as in the previous study \cite{Helfer:2016ljl} 
using the numerically obtained axion star profile for an $m^2\phi^2$ potential 
\cite{Alcubierre:2003sx, Urena-Lopez:2002ptf, Urena-Lopez:2001zjo}. We choose 
the initial hypersurface such that $\phi=0$ and hence $V(\phi)=0$ everywhere. 
The Hamiltonian constraint is thereby satisfied for both the $V=m^2\phi^2$ and 
the axion potential \eqref{eqn:axion_potential} cases. Furthermore, if we 
impose the extrinsic curvature $K_{ij}=0$, the momentum constraint is trivially 
satisfied and all the dynamical information is encoded in the kinetic term of 
the field $\Pi$. The system is evolved in the moving puncture gauge using the 
default 1+log parameters \eqref{eq:default-lapse-params} for the lapse 
evolution equation \eqref{eq:Bona-Masso-slicing} with the exception of $a_1 = 
0$ and the default Gamma-driver shift parameters for 
\eqref{eq:Gamma-driver-1} and \eqref{eq:Gamma-driver-2} with $\eta = \mu$.

\subsubsection{Tagging criteria}
\label{sec:osc_tag}

We demonstrate the suitability of two different tagging methods for tracking the 
axion star evolution in both the stable and unstable cases. In the case of 
collapse to a BH, for both tagging methods, the threshold $\tau_R$ must be 
chosen such that the apparent horizon is covered entirely by the finest 
refinement level, as discussed in \sref{sec:black-hole-tagging}. In our 
case, we choose a maximum refinement level $l_{\rm max} = 8$, which covers up to 
$\mu r \geq 0.5 > \mu M_{\rm ADM}$.

The first tagging criterion we consider is based on physical quantities in the 
simulation, as outlined in Section \ref{sec:physics-tagging}. We choose the 
absolute sum of the terms in the Hamiltonian constraint $\mathcal 
H_{\text{abs}}$ \eqref{eq:abs-Hamiltoniaon-constraint}, setting the criterion 
\begin{equation}\label{eq:Habs}
    C(\mathbf{i}) \equiv \mathcal{H}_{\text{abs}}
\end{equation}
in the tagging indicator function (\ref{eq:tagging-switch}). 

We also show the efficacy of truncation error tagging, outlined in section 
\ref{sec:truncation-tagging}. We choose the variables $f$ as defined in 
(\ref{eq:trunc_error}) to be $\chi, K, \phi$ and $\pi$, as these capture the 
information in the Hamiltonian constraint $\mathcal{H}$. We use the tagging 
criterion \eqref{eq:weighted-truncation-error}, explicitly
\begin{equation}
     C(\mathbf{i}) = \sqrt{(\tau_{l,\chi}(\mathbf{i}))^2 + 
(\tau_{l,K}(\mathbf{i}))^2 + (\tau_{l,\phi}(\mathbf{i}))^2 + 
(\tau_{l,\pi}(\mathbf{i}))^2}\,,
     \label{eq:truncation_error}
\end{equation}
where $\tau_{l,f}(\mathbf{i})$ is defined by \eqref{eq:trunc_error} and we have 
set the normalizing factor $L_f=1$ for all $f$.

\subsubsection{Diagnostics and convergence testing}
\label{sec:osc_diag}

We perform convergence tests using several key physical quantities from the 
evolution. The first quantity is the $L^2$ norm of the Hamiltonian constraint 
violations \eqref{eq:Ham-constraint} $||\mathcal H||_2$ over a coordinate volume 
$\mathcal{V}$
\begin{equation}\label{eq:Ham-2-norm}
    ||\mathcal{H}||_2 = \sqrt{\int_{\mathcal{V}}\rmd^3x\,\mathcal{H}^2}\,.
\end{equation}
For the case of a stable axion star, in order to exclude the constraint 
violation at the outer boundaries, we choose $\mathcal{V}$ to be the volume 
enclosed by a sphere of fixed coordinate radius $r_{\mathrm{out}}$, 
$B_{r_{\mathrm{out}}}$, with a center that coincides with that of the star:
\begin{equation}
    \mathcal{V}=B_{r_{\mathrm{out}}}.
\end{equation}
In the case of an unstable axion star that collapses to a BH, we furthermore 
excise the volume enclosed by a smaller sphere of fixed coordinate radius 
$r_{\mathrm{in}}<r_{\mathrm{out}}$ with the same center in order to exclude the 
constraint violations near the puncture that arises after the collapse. The 
radius $r_{\mathrm{in}}$ is chosen such that the sphere will lie within the 
apparent horizon once it is formed. This means that
\begin{equation}
    \mathcal{V}=B_{r_{\mathrm{out}}}\setminus B_{r_{\mathrm{in}}}.
\end{equation}

For the stable axion star, we also test convergence using the total 
mass of the matter content $M_{\text{mat}}$ within the same volume
\begin{align}
    M_{\text{mat}} = -\int_{\mathcal{V}}\rmd^3x ~ \sqrt{\gamma}\, T_0^0 = 
\int_{\mathcal{V}}{\rmd^3x\,\sqrt{\gamma}\left(\alpha\rho - 
\beta_kS^k\right)}\,,
    \label{eq:M_mat}
\end{align}
where $\rho$ and $S^i$ are defined by \eqref{eq:decomposed-EM-tensor}, 
$\gamma = \det{(\gamma_{ij})}$ and $\alpha$ and $\beta_i$ are the lapse and 
shift as defined by \eqref{eq:adaptedcoords}. Further details can be found 
in \cite{Clough:2021qlv}. 

\subsection{Results}
\label{sec:scalarstars}

We have performed simulations of axion stars at different resolutions for each 
configuration in table \ref{tab:grchombo-grids-oscillaton}. We evolve with 
\grchombo and investigate two different tagging criteria: refinement using (i) 
the Hamiltonian constraint $\mathcal{H}_{\text{abs}}$ \eqref{eq:Habs} and (ii) 
the truncation error of the variables $\chi$, $K$, $\phi$ and $\pi$ as defined 
by \eqref{eq:truncation_error}. 

As outlined in \sref{sec:physics-tagging}, for the stable axion star 
configuration ($f_a=1.0$), we expect Hamiltonian constraint tagging to generate 
stable refinement levels. We set $\tau_R$ such that we obtain an appropriate 
initial grid structure, which we choose to have a maximum refinement level 
$l_{\rm max}=3$ with refinement concentrated on the axion star.

\begin{figure}[t]
    \centering
    \includegraphics[width=0.7\columnwidth]{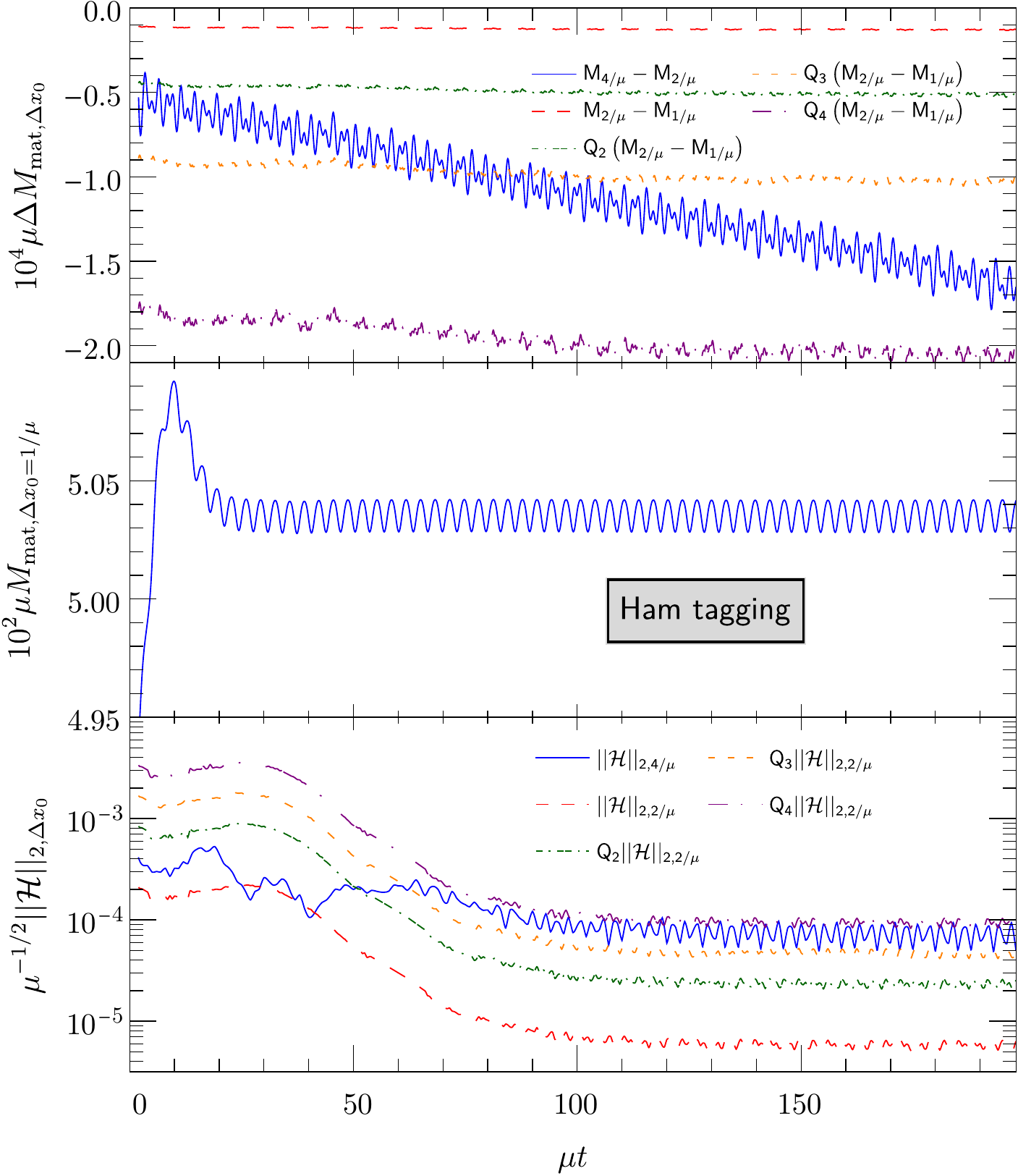}
    \caption{Convergence plots for the stable axion star configuration ($f_a = 
    1.0$) with Hamiltonian constraint tagging \eqref{eq:Habs} and grid 
    configurations given in table \ref{tab:grchombo-grids-oscillaton}. The top  
     panel shows the difference in the calculated matter mass $M_{\text{mat}}$   
     \eqref{eq:M_mat} within a sphere of radius $\mu r_{\text{out}}=25$ between 
    the resolutions with rescalings according to second, third and fourth order 
    convergence. For reference, in the middle panel, we show $M_{\text{mat}}$ 
    for the highest resolution simulation (with $\mu\Delta x_0 = 1$). In the 
    bottom panel, we plot the $L^2$ norm of the Hamiltonian constraint 
    $||\mathcal{H}||_2$ \eqref{eq:Ham-2-norm} for the two lower resolution 
    simulations ($\mu\Delta x_0 =4,2$) in addition to rescalings according to 
    second, third and fourth order convergence. We omit the corresponding plots 
    for the simulations with truncation error tagging 
    \eqref{eq:truncation_error} as they are qualitatively very similar.
    }
    \label{fig:osc-convergence-stable}
\end{figure}

The middle panel of figure \ref{fig:osc-convergence-stable} shows $M_{\rm mat}$ 
for the finest grid configuration for a stable axion star in table 
\ref{tab:grchombo-grids-oscillaton}, where $M_{\rm mat}$ is calculated as 
defined in \eqref{eq:M_mat} for a coordinate sphere with radius $\mu r_{\rm out} 
= 25$. We observe some initial gauge evolution of $M_{\rm mat}$ due to the 
transition from the initial polar-areal gauge to puncture gauge, with subsequent 
regular oscillations over time, which are physical. Given that $M_{\rm mat}$ 
includes only matter contributions and the mass of an axion star is 
approximately constant (i.e. the total flux out of the outer spherical surface 
is zero), these oscillations indicate the transfer of energy between curvature 
and matter terms as discussed in \sref{sec:osc_setup}.

The upper panel of figure \ref{fig:osc-convergence-stable} shows the difference 
in mass $\Delta M_{\rm mat}$ for the simulations in table 
\ref{tab:grchombo-grids-oscillaton}. This demonstrates convergence between third 
and fourth order in $M_{\rm mat}$ at late times, but second order near the 
beginning of the simulation. This agrees with expectations; the initial data 
used is accurate to second order, and this error dominates at early times, with 
the fourth order convergence related to the evolution scheme only being 
recovered at later times. Some error is also introduced by the interpolation of 
the initial conditions onto the grid, which is first order (but with a high 
spatial resolution in the numerical solution so this is subdominant). By 
comparing the highest resolution simulation with a Richardson extrapolation, we 
obtain a discretization error estimate of $\Delta M_{\rm mat}/M_{\rm mat} 
\lesssim 4\times 10^{-5}$ at late times (using third order extrapolation).

The lower panel of figure \ref{fig:osc-convergence-stable} shows the $L^2$ norm 
of the Hamiltonian constraint violations $||\mathcal H||_2$ for the same 
simulations. Again, we measure between third and fourth order convergence at 
late times, and between first and second order initially. We obtain an error 
measure at late times of $\sqrt{M_{\rm ADM}}\,\Delta||\mathcal H||_2 / 16\pi 
\lesssim 8\times10^{-8}$, where we have normalized with the ADM mass to create a 
dimensionless measure of the spurious energy density.

We perform the same analysis of $M_{\rm mat}$ and $||\mathcal H||_2$ for the 
stable axion star using truncation error tagging with the parameters in table 
\ref{tab:grchombo-grids-oscillaton}. We obtain a very similar grid structure and 
evolution behaviour to Hamiltonian tagging, with the same convergence and error 
estimates, demonstrating that both methods can achieve equivalent, accurate 
results.

\begin{figure}[t]
    \centering
    \includegraphics[width=0.7\columnwidth]{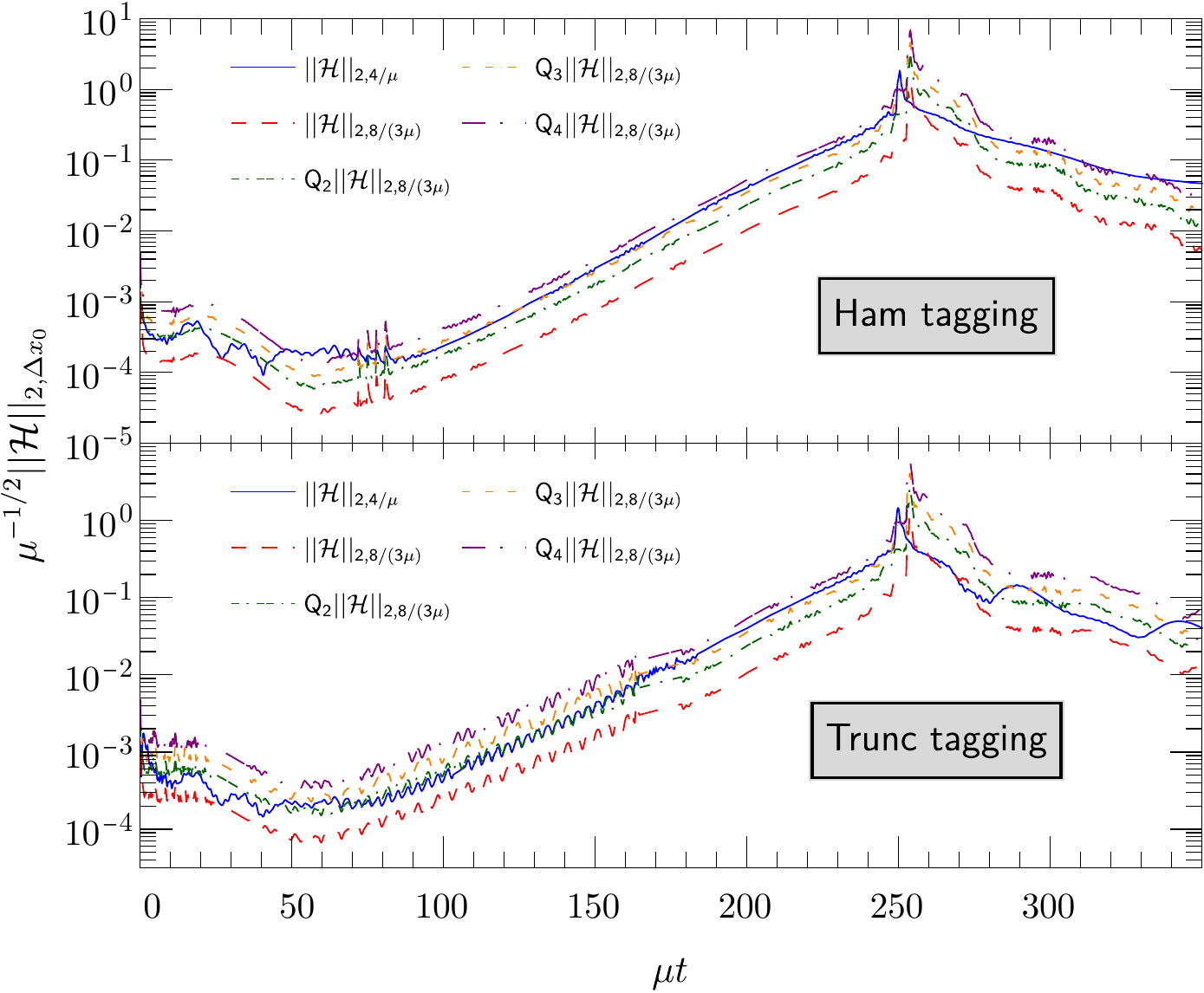}
    \caption{Convergence plots of the $L^2$ norm of the Hamiltonian constraint 
    \eqref{eq:Ham-2-norm} contained between spheres of radius $\mu 
    r_{\text{in}}=0.5$ and $\mu r_{\text{out}} = 25$ for the unstable axion 
    star configuration ($f_a = 0.05$) with both Hamiltonian constraint tagging
    \eqref{eq:Habs} (top panel) and truncation error tagging 
    \eqref{eq:truncation_error} (bottom panel). The grid configurations are 
    provided in table \ref{tab:grchombo-grids-oscillaton} and we additionally 
    plot rescalings according to second, third and fourth order convergence. 
    }
    \label{fig:osc-convergence-collapse}
\end{figure}

For the unstable axion star configuration ($f_a=0.05$), we perform convergence 
testing on $||\mathcal{H}||_2$ within the spatial volume with $0.5 < \mu r < 
25$, excising the region $\mu r < 0.5$ where a BH is formed. We use the 
parameters given in table \ref{tab:grchombo-grids-oscillaton}. We do not 
perform a convergence test of $M_{\rm mat}$, as the quantity oscillates with a 
high frequency about a mean value that rapidly decreases to zero around the 
collapse, making such an analysis impractical.

The top panel of figure \ref{fig:osc-convergence-collapse} shows the 
convergence analysis of $||\mathcal{H}||_2$ with Hamiltonian constraint tagging. 
We observe approximately third order convergence prior to collapse to a BH, 
then approximately fourth order convergence at late time. We obtain a maximum 
error measure on the finer grid of $\sqrt{M_{\rm ADM}} \Delta||\mathcal{H}||_2 
/ 16\pi \lesssim 0.021$, with this value occurring approximately at the 
collapse. 

For the same configuration with truncation error tagging in the lower panel of 
figure \ref{fig:osc-convergence-collapse}, we obtain similar convergence prior 
to the collapse with a maximum error measure $\sqrt{M_{\rm ADM}} 
\Delta||\mathcal{H}||_2 / 16\pi \lesssim 0.015$. The convergence after the 
collapse is lower: between first and third order. In general we see that the 
rapid regridding that occurs during a collapse (often triggered at different 
times at the different resolutions) introduces errors which can reduce the 
convergence order. This illustrates one of the main challenges of AMR, which is 
to obtain good convergence in highly dynamical regimes.

\section{Conclusion}

In this work, we have presented a detailed discussion of the use of fully 
adaptive mesh refinement (AMR) in numerical relativity simulations with the 
{\sc GRChombo} code. To avoid confusion, we first summarize how the term 
``fully adaptive'' is meant to distinguish AMR from the common (and often 
highly successful) box-in-a-box approach. This distinction
consists in two main features. First, we use the term AMR in the sense that it 
allows for refined regions of essentially arbitrary shape. Second, it 
identifies regions for refinement based on a point-by-point interpretation of 
one or more user-specifiable functions of grid variables. Of course, a region 
of arbitrary shape will inevitably be approximated by a large number of boxes 
on Cartesian grids; the distinction from a box-in-a-box approach therefore 
consists in the {\it large number} of boxes used in AMR.
Likewise, every box-in-a-box approach will ultimately base its dynamic 
regridding on some function of the evolved grid variables, as the apparent 
horizon. The key feature of AMR is the {\it pointwise} evaluation of grid 
variables or their derived quantities.

We have laid the foundations for our study in sections \ref{sec:grchombo} and 
\ref{sec:amr-techniques} with a comprehensive summary of the formulation of the 
Einstein equations and the AMR infrastructure of {\sc GRChombo}. In short, we 
employ the CCZ4 equations (\ref{eq:CCZ4-chi})-(\ref{eq:CCZ4-Gamma}), on a 
Cartesian mesh with a user-specified number of refinement levels with sixth or 
fourth-order spatial discretization and fourth-order Runge-Kutta time stepping 
according to the method of lines. The tagging of grid points for refinement and 
the corresponding regridding is performed according to the Berger-Rigoutsos AMR 
algorithm summarized in \sref{sec:BR-AMR} and figure 
\ref{fig:berger-rigoutsos-partitioning}.

The advantages of AMR based simulations over the simpler box-in-a-box structure 
evidently arise from its capability to flexibly adapt to essentially any 
changes in the shape or structure of the physical system under consideration. 
These advantages, however, do not come without new challenges; the 
identification of these challenges and the development of tools to overcome 
them are the main result of our work.

The first and most elementary result of our study is the (hardly surprising) 
observation that there exist no ``one size fits all'' criterion for refinement 
that automatically handles all possible physical systems. Many of the 
challenges, however, can be effectively addressed with a combination of a small 
number of criteria for tagging and refining regions of the domain. We summarize 
these challenges and techniques as follows. 
\begin{enumerate}[label=(\roman*)]
\item
  In AMR it is more difficult to test for (and obtain)
  convergence, because of the loss
  of direct control over the resolution in a given region
  of spacetime. While the refinement in
  AMR is every bit as deterministic as it is in a box-in-a-box
  approach, the complexity of the underlying algorithm makes
  it practically impossible for a user to predict when, if and
  where refinement will take place. Consider for example
  the convergence analysis of a simulation using the
  truncation-error based tagging criterion of
  (\ref{eq:trunc_error}); in some regions of the spacetime
  a low resolution
  run may encounter a sufficiently large truncation error
  to trigger refinement
  whereas a higher-resolution run will not.
  To counteract this effect, one may adjust the tagging threshold
  in anticipation of the reduction in the truncation error,
  but some experimentation is often necessary because
  different ingredients of the code have different orders of accuracy.
  Additionally, one may enforce refinement using
  a priori knowledge, as for example, through enforced
  tagging around the spheres of wave extraction.
  \change{}{An alternative approach would be to record the the grid 
  structure over time for one simulation (e.g. the lowest resolution run) 
  and then ``replay'' this grid structure (or as close as possible to it)
  for simulations at different resolutions as is done for the \textsc{Had}
  code \cite{Liebling:2004nr}.
  }
 \item
  A further challenge arises from the
  use of too many refinement regions/boundaries
  over a small volume in spacetime. The interpolation at
  refinement boundaries is prone to generating small levels
  of high-frequency
  numerical noise that may bounce off neighbouring boundaries
  if these are too close in space (or time). An effective
  way to handle this problem is the use of buffer zones in space
  and to avoid unnecessarily frequent regridding. 
  %
  \item
  In the case of BH simulations, we often observe
  a degradation of numerical accuracy when refinement levels cross or even
  exist close to the apparent horizon.
  This typically manifests itself as an unphysical drift in the horizon area 
  and, in the case of binaries, a loss of phase accuracy and/or a drift in the 
  BH trajectory. These problems can be cured by enforced tagging of all grid 
  points inside the apparent horizon. In practice, we add an additional buffer
  zone to ensure all refinement boundaries are sufficiently
  far away from the apparent horizon(s). 
  \item
  The Berger-Rigoutsos algorithm detailed in \sref{sec:BR-AMR}
  does not treat the $x$, $y$ and $z$ direction
  on exactly equal footing; the
  partitioning algorithm
  (cf.~figure \ref{fig:berger-rigoutsos-partitioning})
  inevitably handles the coordinate directions in a specific order. This
  can lead to asymmetries in the refined grids even when
  the underlying spacetime region is symmetric.
  In some simulations of BHs, we noticed this to cause a
  loss in accuracy.
  A simple way to overcome this problem is to enforce a boxlike
  structure around BHs. 
  \item
  A single tagging variable (or one of its 
  spatial derivatives) 
  may not always be suitable to achieve
  appropriate refinement 
  throughout the course of an entire simulation;
  for example, this may be due to gauge dependence or dramatic
  changes in the dynamics of the physical evolution.
  {\sc GRChombo} allows for tagging regions based on 
  arbitrary functions of
  {\it multiple} variables and their derivatives
  to overcome problems of this kind.
\end{enumerate}
In order to avoid the difficulties listed here,
we often combine two or more
tagging criteria. The efficacy of this approach
is demonstrated in sections \ref{sec:bh} and \ref{sec:osc}
where we present in detail AMR simulations of inspiraling
BH binaries and stable as well as collapsing axion stars.
By comparing the BH simulations with those from the
box-in-a-box based {\sc Lean} code, we demonstrate
that with an appropriate choice of tagging criteria,
AMR simulations reach the same accuracy and convergence
as state-of-the-art BH binary codes using Cartesian grids.
While AMR does not directly bestow major benefits on the
modeling of vacuum BH binaries\change{}{ (and is typically more
computationally expensive)}, it offers greater
flexibility in generalizing these to BH
spacetimes with scalar or vector fields, other forms
of matter, or BHs of nearly fractal shape that can form
in higher-dimensional collisions \cite{Andrade:2020dgc}.
The simulations of rapidly oscillating or gravitationally
collapsing scalar fields demonstrate {\sc GRChombo}'s
capacity to evolve highly compact and dynamic matter
configurations of this type.

%
Finally, we note the potential of AMR for hydrodynamic simulations.
However, high-resolution shock capturing methods present qualitatively
new challenges for AMR and we leave the investigation of this
topic for future work.


\ack

\censor{We thank the rest of the GRChombo collaboration
(\url{www.grchombo.org})}
\xblackout{
for their support and code development work, and for helpful discussions regarding 
their use of AMR.}
\xblackout{
MR acknowledges support from a Science and Technology Facilities Council (STFC) 
studentship.
KC acknowledges funding from the European Research Council (ERC) under the 
European Unions Horizon 2020 research and innovation programme (grant agreement 
No 693024), and an STFC Ernest Rutherford Fellowship. AD is supported by a 
Junior Research Fellowship (JRF) at Homerton College, University of Cambridge.
PF is supported by the European Research Council Grant No.~ERC-2014-StG
639022-NewNGR, and by a Royal Society University Research Fellowship Grants 
No.~UF140319, RGF\textbackslash EA\textbackslash180260, 
URF\textbackslash R\textbackslash 201026 and RF\textbackslash 
ERE\textbackslash 210291.
JCA acknowledges funding from the European Research Council (ERC) under the 
European Unions Horizon 2020 research and innovation programme (grant agreement 
No 693024), from the Beecroft Trust and from The Queen's College via an 
extraordinary Junior Research Fellowship (eJRF). TF is supported by a Royal 
Society Enhancement Award (Grant No.~
RGF{\textbackslash}EA{\textbackslash}180260). TH is supported by NSF Grants 
Nos.~PHY-1912550, AST-2006538, PHY-090003 and PHY-20043, and NASA Grants Nos.~%
17-ATP17-0225, 19-ATP19-0051 and 20-LPS20-0011.
JLR and US are supported by STFC Research Grant
No.~ST/V005669/1; US is supported by the
H2020-ERC-2014-CoG Grant ``MaGRaTh'' No.~646597.}

\xblackout{
The simulations presented in this paper used DiRAC resources under the projects 
ACSP218, ACTP186 and ACTP238. The work was performed using the Cambridge Service for Data 
Driven Discovery (CSD3), part of which is operated by the University of 
Cambridge Research Computing on behalf of the STFC DiRAC HPC Facility 
(www.dirac.ac.uk). The DiRAC component of CSD3 was funded by BEIS capital 
funding via STFC capital grants ST/P002307/1 and ST/R002452/1 and STFC 
operations grant ST/R00689X/1. In addition we have used the DiRAC at Durham 
facility managed by the Institute for Computational Cosmology on behalf of the 
STFC DiRAC HPC Facility (www.dirac.ac.uk). The equipment was funded by BEIS 
capital funding via STFC capital grants ST/P002293/1 and ST/R002371/1, Durham 
University and STFC operations grant ST/R000832/1. DiRAC is part of the National 
e-Infrastructure.
XSEDE resources under Grant No.~PHY-090003 were used
on
the San Diego Supercomputing Center's clusters Comet and Expanse and
the Texas Advanced Supercomputing Center's (TACC) Stampede2. Furthermore,
we acknowledge the use of HPC resources on the TACC Frontera cluster} 
\censor{\cite{Stanzione:2020}.}
\xblackout{PRACE resources under Grant Number 2020225359 were used on the 
GCS Supercomputer 
JUWELS at the} \censor{J\"ulich} 
\xblackout{Supercomputing Centre (JCS) through the John von Neumann 
Institute for Computing (NIC), funded by the Gauss Centre for Supercomputing 
e.V.}
\censor{(\url{www.gauss-centre.eu}).} 
\xblackout{
The Fawcett supercomputer at the Department of 
Applied Mathematics and Theoretical Physics (DAMTP), University of Cambridge was 
also used, funded by STFC Consolidated Grant ST/P000673/1.}

\appendix


\section{Calculation of the Weyl Scalar\texorpdfstring{ $\Psi_4$}{}}
\label{sec:psi4-appendix}

The Weyl tensor \cite{Weyl:1918pdp} in four spacetime dimensions is defined by
\begin{equation}
    C_{\mu\nu\rho\sigma} := {}^{(4)}R_{\mu\nu\rho\sigma} - 
    \left(g_{\mu[\rho}\,{}^{(4)}R_{\sigma]\nu} -
    g_{\nu[\rho}\,{}^{(4)}R_{\sigma]\mu}\right)
    + \frac{1}{3}g_{\mu[\rho}g_{\sigma]\nu}\,{}^{(4)}R.
\end{equation}
It is completely determined by its electric and magnetic parts 
\cite{Stephani:2004ud}
\begin{align}
    E_{\mu\nu} &:= n^\alpha n^\beta C_{\alpha\mu\beta\nu},\\
    B_{\mu\nu} &:= n^\alpha n^\beta (\ast C)_{\alpha\mu\beta\nu},
\end{align}
where the dual Weyl tensor $(\ast C)_{\mu\nu\rho\sigma}$ is given by
\begin{equation}
    (\ast C)_{\mu\nu\rho\sigma} := 
    \frac{1}{2}\epsilon^{\alpha\beta}_{\phantom{\alpha\beta}\rho\sigma}
    C_{ \mu\nu\alpha\beta},
\end{equation}
and $\epsilon_{\mu\nu\rho\sigma}$ is the volume form.
Because of the symmetries of the Weyl tensor, the electric and magnetic parts 
are symmetric, trace-free and purely spatial.

In the Newman-Penrose formalism \cite{Newman:1961qr}, one introduces a complex 
null tetrad $(l^\mu,k^\mu,m^\mu,\bar{m}^\mu)$, where we follow the notation of 
\cite{Alcubierre:2008co} in order to avoid confusion with the normal 
$n^{\mu}$ to the foliation \eqref{eq:normal}. The Newman-Penrose, or Weyl, 
scalar $\Psi_4$ is defined by
\begin{equation}
    \Psi_4 
    := 
    C_{\alpha\beta\gamma\delta} k^\alpha \bar{m}^\beta k^\gamma \bar{m}^\delta,
\end{equation}
which can be shown to reduce to \cite{Alcubierre:2008co}
\begin{equation}
    \Psi_4 := (E_{ij} - \iu B_{ij})\bar{m}^i\bar{m}^j.
\end{equation}

We use the approach described in step (a) of section V A of 
\cite{Baker:2001sf} to construct a null tetrad with the inner products
\begin{equation}
    -l_\alpha k^\alpha = m_\alpha \bar{m}^\alpha = 1,
\end{equation}
and all others vanishing. Following \cite{Bruegmann:2006ulg,Fiske:2005fx}, 
we omit the null rotations in order to bring the tetrad into a quasi-Kinnersley 
form (step (b) in section V A of \cite{Baker:2001sf}). The expressions for 
the electric and magnetic parts of the Weyl scalar in the 3+1 Z4 formulation are
\begin{align}
    E_{ij} &= \left[R_{ij} - K_i^{\phantom{i}m}K_{jm} + K_{ij}(K - 
    \Theta)-4\pi S_{ij} + D_{(i}\Theta_{j)}\right]^{\text{TF}},
    \\
    B_{ij} &= \epsilon_{mn(i}D^mK_{j)}^{\phantom{j)}n},
\end{align}
where $\epsilon_{ijk}=n^\alpha\epsilon_{\alpha ijk}$ is the volume form on the 
hypersurface. Note that unlike the usual 3+1 expressions, for example 
(8.3.15)-(8.3.16) in \cite{Alcubierre:2008co}, these expressions are 
manifestly symmetric and trace-free.


\section{Spatial Derivative Stencils}
\label{sec:stencils}

We use the formulae in \cite{Zlochower:2005bj} for the fourth order 
stencils. Using the conventional notation for finite differences where,
\begin{equation}
    F_i = F|_{x=x_i},\qquad F_{i,j} = F|_{x=x_i,y=y_j},
\end{equation}
and $x_i$ and $y_i$ are coordinates of the discrete points on a uniform grid, 
the centered stencils are
\begin{align}
    \partial_x F &= \frac{1}{12h}\left(F_{i-2} - 8 F_{i-1} 
    + 8 F_{i+1} - F_{i+2}\right),
    \\
    \partial^2_x F &= \frac{1}{12h^2}\left(-F_{i-2} + 16 F_{i-1} - 30 F_i      
    + 16 F_{i+1} - F_{i+2}\right),
    \\
    \begin{split}
        \partial^2_{xy}F &= \frac{1}{144h^2}\left(F_{i-2,j-2} - 8 F_{i-2,j-1}
        + 8 F_{i-2,j+1} - F_{i-2,j+2} - 8F_{i-1,j-2}\right.
        \\ 
        &\quad + 64F_{i-1,j-1} - 64F_{i-1,j+1} + 8F_{i-1,j+2} + 8F_{i+1,j-2}
        - 64F_{i+1, j-1}
        \\
        &\quad + 64F_{i+1,j+1} - 8F_{i+1,j+2} - F_{i+2,j-2} + 8F_{i+2,j-1} 
        - 8F_{i+2,j+1}
        \\
        &\quad\left.+ F_{i+2,j+2}\right),
    \end{split}
\end{align}
and, for the advection term, the lopsided stencils are
\begin{align}
    \partial_x F &= \frac{1}{12h}\left(-3 F_{i-1} - 10F_i + 18F_{i+1}
    - 6F_{i+2} + F_{i+3}\right)\text{ if }\beta^x > 0,
    \\
    \partial_x F &= \frac{1}{12h}\left(- F_{i-3} + 6F_{i-2} - 18F_{i-1}
    + 10F_i + 3F_{i+1}\right)\text{ if }\beta^x \leq 0.
\end{align}
We follow \cite{Husa:2007hp} for the sixth order stencils. The centered stencils 
are
\begin{align}
    \partial_x F &= \frac{1}{60h}\left(-F_{i-3} + 9F_{i-2} - 
    45F_{i-1} + 45F_{i+1} - 9 F_{i+2} + F_{i+3}\right),
    \\
    \begin{split}
        \partial^2_x F &= \frac{1}{180h^2}\left(2 F_{i-3} - 27F_{i-2} 
        + 270F_{i-1} - 490F_i + 270F_{i+1} - 27F_{i+2} 
        \right.\\&\quad\left.
        + 2F_{i+3}\right),
    \end{split}
    \\
    \begin{split}
        \partial^2_{xy} F &= \frac{1}{3600h}\left(F_{i-3.j-3} - 9F_{i-3,j-2}
        + 45F_{i-3,j-1} - 45F_{i-3.j+1}\right.
        \\
        &\quad  + 9 F_{i-3,j+2} - F_{i-3,j+3} - 9F_{i-2,j-3} + 81F_{i-2,j-2}
        \\
        &\quad - 405F_{i-2,j-1} + 405F_{i-2,j+1} - 81F_{i-2,j+2} + 9F_{i-2,j+3}
        \\
        &\quad + 45F_{i-1,j-3} - 405F_{i-1,j-2} + 2025F_{i-1,j-1}
        - 2025F_{i-1,j+1} 
        \\
        &\quad + 405F_{i-1,j+2} - 45F_{i-1,j+3} - 45F_{i+1,j-3} + 405F_{i+1,j-2}
        \\
        &\quad - 2025F_{i+1,j-1} + 2025F_{i+1,j+1} - 405F_{i+1,j+2} 
        + 45F_{i+1,j+3}
        \\
        &\quad + 9F_{i+2,j-3} - 81F_{i+2,j-2} + 405F_{i+2,j-1} - 405F_{i+2,j+1}
        \\
        &\quad + 81F_{i+2,j+2} - 9F_{i+2,j+3} - F_{i+3,j-3} + 9F_{i+3,j-2} 
        \\
        &\quad\left. - 45F_{i+3,j-1} + 45F_{i+3,j+1} - 9F_{i+3,j+2} 
        + F_{i+3,j+3}\right),
    \end{split}
\end{align}
and, for the advection terms, the lopsided stencils are
\begin{align}
    \begin{split}
        \partial_x F &= \frac{1}{60h}\left( 2F_{i-2} - 24F_{i-1} - 35F_i 
        + 80F_{i+1} - 30 F_{i+2} + 8F_{i+3} 
        \right. \\&\quad\left.
         - F_{i+4}\right)\text{ if }\beta^x > 0,
    \end{split}
    \\
    \begin{split}
        \partial_x F &= \frac{1}{60h}\left(F_{i-4} - 8F_{i-3} + 30 F_{i-2}
        - 80F_{i-1} + 35F_i + 24F_{i+1} 
        \right. \\ &\quad\left. 
        - 2F_{i+2}\right)\text{ if }\beta^x < 0.
    \end{split}
\end{align}


\section{Approximate horizon locations as a tagging criteria}
\label{sec:chihorizon}

\begin{figure}[t]
    \centering
    \includegraphics[width=\textwidth]{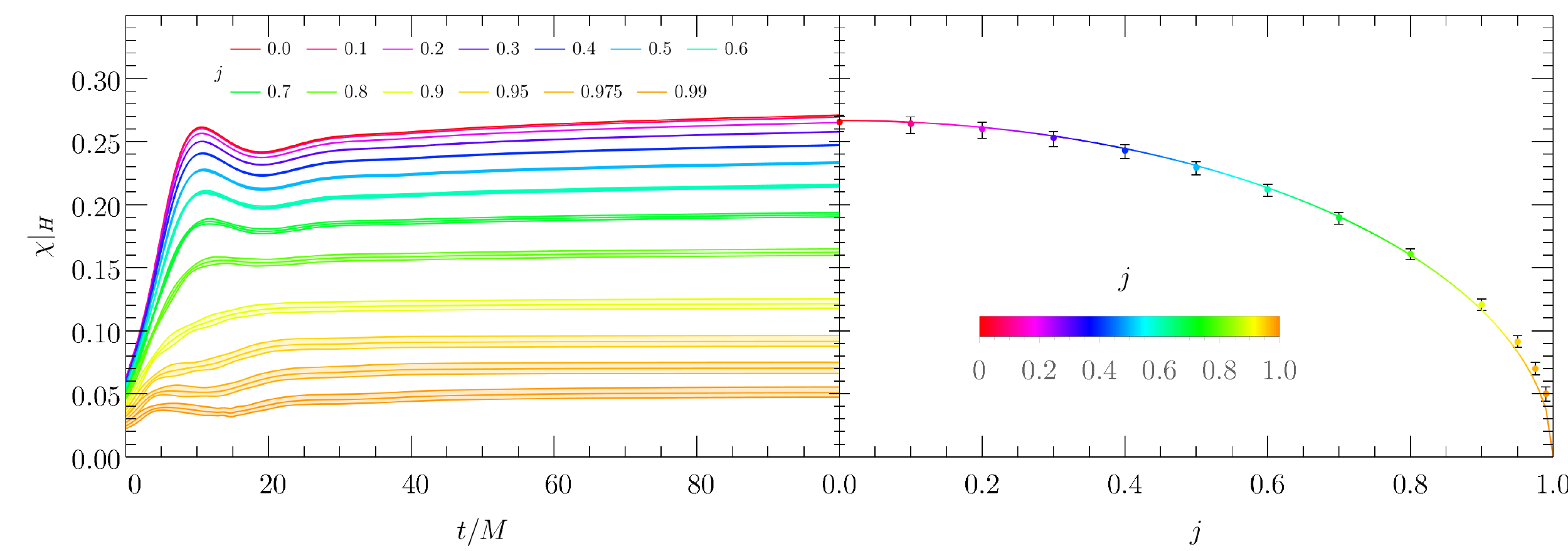}
    \caption{Plots illustrating the dependence of the value of the conformal 
    factor $\chi$ on the apparent horizon surface $H$ in the moving puncture 
    gauge \eqref{eq:Bona-Masso-slicing}, 
    \eqref{eq:Gamma-driver-1}-\eqref{eq:Gamma-driver-2} for different values of 
    the dimensionless spin $j$. For all plots, we use the quasi-isotropic
    Kerr initial data described in \sref{sec:initial-data} and the default 
    values of the gauge parameters \eqref{eq:default-lapse-params} with $M\eta = 
    1$. Although we would expect the plots to vary for different gauge 
    parameters (in particular, as $\eta$ is varied), these plots provide a rough 
    rule-of-thumb. The left panel shows the mean value of $\chi$ as a function 
    of time with the error bands around each curve corresponding to the maximum 
    and minimum on $H$. The right panel shows the mean value of $\chi$ over the 
    interval $t/M\in[40,100]$ for each $j$ with the error bars corresponding to 
    the minimum and maximum values of $\chi$ over the same interval. 
    Furthermore, we show a fit of the mean value of $\chi$ against $j$ which 
    takes the form $\langle\chi\rangle|_H\simeq 0.2666\sqrt{1-j^2}$.}
    \label{fig:chitagging}
\end{figure}
As noted in section \ref{sec:amr-techniques}, the use of the horizon location can 
be an essential part of an adaptive mesh scheme for refinement. In particular, 
one does not usually want to put additional refinement within a horizon (where 
effects are unobservable anyway), and should take care to avoid grid boundaries 
overlapping the horizon, since this can lead to instabilities that strongly 
affect the physical results. Whilst using an apparent horizon finder for this is 
a possibility, often a more ``quick and dirty'' scheme using contours of the 
conformal factor can be just as effective, and significantly easier to 
implement. Whilst in principle there is a dependence on simulation and gauge 
parameters (in particular $\eta$ \cite{Bruegmann:2006ulg}), in general the 
approximate values are quite robust. 

The key dependence is on the [dimensionless] spin of the black hole $j$, as 
illustrated in figure \ref{fig:chitagging}, with a good fit obtained from the 
relation 
\begin{equation}
    \langle\chi\rangle|_H = 0.2666\sqrt{1-j^2}.
\end{equation}
One key advantage is that one does not need to know a priori the mass of the BH 
spacetime which forms, and it can be seen that simply using the $j=0$ values 
will give a conservative coverage of the horizon. These types of criteria were 
used extensively in the higher dimensional black ring spacetimes studies in 
\cite{Andrade:2020dgc,Bantilan:2019bvf,Figueras:2017zwa,Figueras:2015hkb} 
and for the investigation into gravitational collapse in a modifed gravity 
theory in \cite{Figueras:2020dzx}.


\section{Parallelization in \grchombo}
\label{sec:parallelization}

Like other numerical relativity codes and, more generally, scientific computing 
codes, \grchombo exploits parallelization at several different levels in order 
to achieve good performance and scaling on modern supercomputers\change{}{ (see 
section 2.4 in \cite{Kunesch:2018jeq} for scaling results)}.

For each AMR level, \grchombo splits the domain into boxes and these boxes are 
shared between processes running on multiple distributed-memory nodes using the 
Message Passing Interface (MPI). In practice, even though the memory is shared 
within a node, we typically still use multiple MPI processes per node in order 
to achieve optimal performance. For example, if a node has $n$ cores, we might 
choose to use between $n/4$-$n/2$ MPI processes per node. At every 
\emph{regrid}, we use a load balancing routine in \textsc{Chombo} in order to 
evenly distribute the boxes. We sort the boxes using a Morton ordering, as this 
minimizes communication by increasing the chance that neighboring boxes are on 
the same or nearby MPI processes.

One of the most common operations in an NR code is looping through all the 
cells/points on the grid, calculating some expression and then storing its 
value in a grid variable. An example is the calculation of the RHS at every RK4 
substep which is often where a code spends a large proportion of its time. 
Within an MPI process, \grchombo uses OpenMP to thread these loops over the $z$ 
and $y$ coordinates of the boxes. For the $x$ direction, \grchombo relies on 
SIMD (single instruction, multiple data)/vector 
intrinsics\footnote{For the x86\_64 architecture, \grchombo 
currently supports SSE2, AVX and AVX-512 instructions. } in order to utilize 
the full vector-width of the targeted architecture. We use intrinsics because 
the complexity of the CCZ4 equations \eqref{eq:CCZ4-chi}-\eqref{eq:CCZ4-Gamma} 
means that compilers will usually fail to auto-vectorize these loops. The main 
disadvantage of using SIMD intrinsics is that they are complex and difficult to 
implement properly. In order to hide the technical implementation from users, 
many NR codes rely on code-generation scripts to convert more familiar 
Mathematica/Python expressions to optimized and vectorized Fortran/C/C++ code, 
for example, \textsc{Kranc} \cite{Husa:2004ip} and \textsc{NRPy+} 
\cite{Ruchlin:2017com} . \grchombo takes a different approach, keeping the 
programming at the lower level but relying on C++14 templates to provide a 
somewhat more user-friendly interface for writing optimized code. Vectorized 
expressions can be enforced by replacing the C++ type \texttt{double} with a 
template type \texttt{data\_t}, which represents a vector of values of the 
variables on the grid of arbitrary length (e.g. the value of $\chi$ at the 
points with $x$ index $i_x = 0,1,2,3,\ldots$, and constant $y$ and $z$). In 
order to make this functionality work, the user is required to write their code 
in a \emph{compute class} with a \texttt{compute} member function which can 
then be instantiated as an object and then passed to a loop function which 
calls the \texttt{compute} member function in each vector of cells. Multiple 
compute objects can be combined into a \emph{compute pack} which can then be 
called by the loop function for added efficiency. For a more detailed 
description and examples, see section 2.5 in \cite{Kunesch:2018jeq}.


\bibliographystyle{iopart-num-mod}
\bibliography{bibliography}

\end{document}